\documentclass[%
 reprint,
preprintnumbers,
 amsmath,amssymb,
 aps,
]{revtex4-2}

\usepackage[dvipsnames]{xcolor}
\usepackage{graphicx}
\usepackage{dcolumn}
\usepackage{bm}
\usepackage{hyperref}
\usepackage{multirow}

\definecolor{myblue}{rgb}{0.152941176,0.549019608,0.670588235}
\hypersetup{
    pdftoolbar=true,        
    pdfmenubar=true,        
    pdffitwindow=false,     
    pdfstartview={FitH},    
    pdftitle={Violation of custodial symmetry from the W-boson mass measurement},
    pdfauthor={Ayan Paul and Mauro Valli},
    pdfkeywords={W mass} {custodial symmetry} {SMEFT} {CDF},
    pdfnewwindow=true,
    colorlinks=true,
    linkcolor=myblue,
    citecolor=myblue,
    filecolor=myblue,
    urlcolor=myblue,
    linktocpage=true
}

\def\equationautorefname~#1\null{Eq.\,(#1)\null}
\newcommand{\appendixref}[1]{\hyperref[#1]{appendix~\ref{#1}}}

\begin{document}

\preprint{DESY 22-066}
\preprint{HU-EP-22/16}
\preprint{YITP-SB-2022-15}

\title{Violation of custodial symmetry from W-boson mass measurements}%

\author{Ayan Paul}
\email{ayan.paul@desy.de}
\affiliation{Deutsches Elektronen-Synchrotron DESY, Notkestr. 85, 22607 Hamburg, Germany\\
Institut f\"ur Physik, Humboldt-Universit\"at zu Berlin, D-12489 Berlin, Germany}
 
\author{Mauro Valli}%
 \email{mauro.valli@stonybrook.edu}
\affiliation{%
 C.N. Yang Institute for Theoretical Physics, Stony Brook University, Stony Brook, NY 11794,~USA
}%

\date{\today}

\begin{abstract}
The new measurement of the $W$-boson mass from the CDF collaboration shows a significant tension with the Standard Model prediction. We quantify this discrepancy within a state-of-the-art analysis of electroweak precision data and scrutinize the leading deformations of the Standard Model Effective Field Theory arising at dimension six. We find evidence for a non-zero value of the $T$ parameter, i.e. for a novel source of violation of custodial symmetry, pointing to physics beyond the Standard Model at the 4.5$\sigma$ level. We contextualize the implications of our findings in light of other present anomalies in Particle Physics.
\end{abstract}

\maketitle

\section{Introduction} 
\label{sec:intro}
The CDF collaboration has recently delivered a very precise measurement of the $W$-boson mass, with $M_W=80.434\pm0.009$ GeV, reaching the level of 0.01\% precision~\cite{CDF:2022hxs} superseding the previous CDF measurement~\cite{CDF:2012gpf}. Such a phenomenal result provides an unprecedented probe of the underlying dynamics in the 2-point function of the $SU(2)_{L}$ gauge-boson field, following up on recent dedicated studies from ATLAS, D0, and LHCb collaborations~\cite{ATLAS:2017rzl,D0:2013jba,LHCb:2021bjt}. At a first glance, the Standard Model (SM) expectation of $M_{W}=80.357\pm 4_{inputs}\pm4_{theory}$ GeV~\cite{Zyla:2020zbs} establishes a tension of about 7$\sigma$ with the new CDF II measurement, with the latter leaving room for a possible indirect imprint of New Physics (NP) beyond the SM.

Decades of theoretical development have brought about very precise computations of the electroweak (EW) precision observables (EWPO) which constitute a fundamental testbed for the standard theory, yielding powerful constraints in many scenarios of physics beyond the SM (BSM)~\cite{Sirlin:1980nh,Marciano:1980pb,Djouadi:1987gn,Djouadi:1987di,Kniehl:1989yc,Halzen:1990je,Kniehl:1991gu,Kniehl:1992dx,Barbieri:1992nz,Barbieri:1992dq,Djouadi:1993ss,Fleischer:1993ub,Fleischer:1994cb,Avdeev:1994db,Chetyrkin:1995ix,Chetyrkin:1995js,Degrassi:1996mg,Degrassi:1996ps,Degrassi:1999jd,Freitas:2000gg,vanderBij:2000cg,Freitas:2002ja,Awramik:2002wn,Onishchenko:2002ve,Awramik:2002vu,Awramik:2002wv,Awramik:2003ee,Awramik:2003rn,Faisst:2003px,Dubovyk:2016aqv,Dubovyk:2018rlg}. Of particular significance, the misalignment between charged and neutral EW boson masses offers a deep insight into the SM theory and its possible extensions, being a remnant of the $SU(2)_{L} \times SU(2)_{R}$ custodial symmetry present in the Higgs sector, but broken in the SM by hypercharge and Yukawa couplings~\cite{Barbieri:2007gi}.

Putting the above theoretical considerations next to the very accurate measurement of the $Z$-boson mass of the LEP precision program, it should be clear that any relevant experimental progress on the determination of the $W$-boson mass may represent a pillar for advancement in the field. In this work, motivated by such an opportunity, we explore some of the most evident implications of the recent CDF II measurement of $M_W$.  We will focus, in particular, on how the prediction of $M_{W}$ gets affected in SM extensions where a NP mass gap above the EW scale is present, and $SU(2)_{L} \times U(1)_{Y}$ symmetry can be linearly realized once heavy new dynamics is integrated out. Any source of BSM physics of this sort should be described by means of the widely studied Standard Model Effective Field Theory (SMEFT), see e.g.~\cite{Ellis:2020unq,DeBlas:2019qco,deBlas:2019rxi,Ellis:2018gqa}. Within the simplifying assumption of $U(3)^5$ flavor universality for the NP effects under scrutiny, in the following, we make an attempt to learn about possible new sources of custodial-symmetry breaking supported by current data~\footnote{As recently discussed in Ref.~\cite{Kribs:2020jgn}, a general characterization of UV breaking effects of $SU(2)_{L} \times SU(2)_{R}$ may require some care. In this study, we simply regard these effects as those originally highlighted in Ref.~\cite{Peskin:1990zt} in the limit of zero momentum.}.

In \autoref{sec:dim6} we characterize in greater detail our study of the EW sector in the SMEFT; in \autoref{sec:mod} we describe the strategy of our analysis; in \autoref{sec:lessons} we detail our most important findings, reporting evidence for a new source of custodial-symmetry breaking at the level of $4.5\sigma$; in \autoref{sec:discussion} we briefly discuss some implications regarding other ongoing anomalies in Particle Physics and draw our conclusions.

\section{W-boson mass at dimension six} 
\label{sec:dim6}

It is a well-known fact that the leading deformations of the SM EW sector can be encoded in 10 SMEFT operators at dimension six. However, only 8 independent directions can be constrained using the EWPO, see e.g.~\cite{DeBlas:2019qco,Alasfar:2020mne}. Nevertheless, in the Warsaw basis~\cite{Grzadkowski:2010es}, the independent SMEFT operators that explicitly contribute to the $W$-boson mass are~\cite{Berthier:2015oma}: 
{\allowdisplaybreaks
\begin{eqnarray}
\label{eq:OpeSMEFT}
\mathcal{O}_{HWB} &=& H^\dagger \sigma^a H \, W^a_{\mu\nu}B^{\mu\nu}, \nonumber \\
\mathcal{O}_{HD} &=& \left(H^\dagger D^{\mu} H\right)^*\left(H^\dagger D_{\mu} H\right) , \nonumber\\
\mathcal{O}_{LL} &=& \left(\bar{L}\gamma^\mu L\right)\left(\bar{L}\gamma_\mu L\right) \ , \\
\mathcal{O}_{HL}^{(3)} &=& \left(H^\dagger i \overleftrightarrow{D}^a_\mu H\right)\left(\bar{L}\gamma^\mu\sigma^a L\right) \nonumber \ .
\end{eqnarray}
}. 

Indeed, at the linear level in NP effects, the relative shift, $\delta M_{W}^2/M_{W}^2$, in the SMEFT at dimension six is proportional to the combination~\cite{Bjorn:2016zlr}:
\begin{equation}
\label{eq:WmassSMEFT}
   4 C_{HWB}+\cot{\theta}_W C_{HD} + 2 \tan{\theta}_{W} (2C_{HL}^{(3)}- C_{LL}) \ ,
\end{equation}
with $\theta_{W}$ being the tree-level weak-mixing angle. Given the focus of the present paper on the effects of the SMEFT operators (described in the Warsaw basis~\cite{Grzadkowski:2010es}) that enter in \autoref{eq:WmassSMEFT}, we can restrict here the discussion to four operators of \autoref{eq:OpeSMEFT}, and their Wilson coefficients, $C_{HWB}$, $C_{HD}$, $C_{LL}$ and $C_{HL}^{(3)}$, assuming lepton universality, all of which can be constrained by EWPO data.

It should be noted that $C_{HWB}$, $C_{HD}$ also modifies the $Z$-boson mass while all four of these coefficients affect $\sin^2\theta_W$ and the width of the $W$ boson, while gauge couplings are affected by $C_{HWB}$ and $C_{HL}^{(3)}$~\cite{Berthier:2015oma}. Hence, any modification in the $W$ mass will eventually be correlated with several other EWPO, highlighting the importance of a global analysis. In the following, we will focus our attention on any deviation from the SM sourced only by these four operators and assume that NP will not generate other dimension-six operators that may also characterize the study of precision measurements of $Z,W$ boson couplings to leptons and quarks, see for instance~\cite{Efrati:2015eaa}.

The focus on the set of operators in \autoref{eq:WmassSMEFT} provides a concrete practical ground where to explore oblique NP contributions, namely NP effects affecting the vacuum-polarization diagrams of EW gauge bosons, typically described in terms of $S$, $T$ and $U$ parameters~\cite{Peskin:1990zt,Peskin:1991sw} (or equivalently, in terms of $\epsilon_1$, $\epsilon_2$ and $\epsilon_3$~\cite{Altarelli:1990zd,Altarelli:1991fk,Altarelli:1993sz}). Note that the SMEFT encodes the broad class of theories where $U$ receives corrections at dimension eight or higher~\cite{1991PhLB..265..326G}: To the level of precision of the present analysis, $U = 0$. On the other hand, the commonly noted $S$, $T$ parameters at dimension six in the SMEFT correspond to:
\begin{equation}
    \alpha S = \frac{2\sin2\theta_W}{\sqrt{2}G_F}C_{HWB}\;, \ \alpha T = -\frac{1}{2\sqrt{2}G_F}C_{HD} \ .
    \label{eq:ST}
\end{equation}
Therefore, in our analysis, new sources of  custodial-symmetry violation will be potentially spotted by a non-zero inference of $C_{HD}$. The considerations above shape the strategy for the fits we propose in the next section to study the new $W$-mass measurement.

\section{Modus operandi}
\label{sec:mod}

The core input parameters of our analysis are reported in \autoref{tab:params}. We adopt Gaussian priors for the $Z$-boson mass, $M_{Z}$, the top-quark mass, $m_{t}$, the strong coupling constant, $\alpha_s(M_Z)$, and the hadronic-loop contribution to the electromagnetic coupling constant, $\Delta \alpha_{\rm had}^{(5)}(M_Z)$. In our analysis, we fix the values of $G_{F}$ and $\alpha_{\rm em}(0)$ since their measurements are well beyond per-mille level precision. As for the $W$-boson mass, we use two different reference values in this work:
\begin{eqnarray}
    M_W^{\rm 2021} &=& 80.379 \pm 0.012 \;{\rm GeV}\ , \nonumber\\
    M_W^{\rm 2022} &=& 80.4060 \pm 0.0075 \;{\rm GeV}\ ,
    \label{eq:mw_values}
\end{eqnarray}
where {\em 2021} refers to the previous global average~\cite{Zyla:2020zbs} before the current CDF II measurement and {\em 2022} refers to our naive combination of the 2021 global average and the new CDF II measurement after removing the previous CDF II measurement. This is done since the current CDF II measurement subsumes all previous CDF II measurements~\cite{CDF:2022hxs} as all the data used in the previous analyses~\cite{CDF:2007cmy,CDF:2012gpf} are also included in the current CDF II measurement.

Using \texttt{HEPfit}~\cite{DeBlas:2019ehy}, we perform a Bayesian fit of EWPO data, the values for which can be found in~\cite{deBlas:2021wap}, for the SM and in several NP scenarios characterized by different combinations of non-vanishing Wilson coefficients. To perform model comparison of different scenarios, we compute the \textit{Information Criterion} (IC) \cite{IC}: 
\begin{equation}\label{eq:IC}
   IC \equiv -2 \overline{\log \mathcal{L}} \, + \, 4 \sigma^{2}_{\log \mathcal{L}} \,,
\end{equation}
where the first and second terms represent the mean and variance of the log-likelihood posterior distribution, respectively. The first term measures the quality of the fit, while the second one counts effective degrees of freedom and thus penalizes models with a larger number of parameters. Models with lower IC should be preferred according to the canonical scale of evidence of Ref.~\cite{BayesFactors}, related in this context to (positive) IC differences.

\def\arraystretch{1.1}
\begin{table}[t]
    \centering
    \begin{tabular}{c|r@{ $\pm$ }l}
    \toprule
         Main Inputs                    &  Mean & Std       \\
         \hline
         $\alpha_s(M_Z)$                & 0.1177 & 0.0010   \\
         $\Delta\alpha^{(5)}_{had}(M_Z)$      & 0.02766 & 0.00010 \\
         $m_{top}$ [GeV]                & 172.58 & 1.00     \\
         $m_{H}$ [GeV]                  & 125.25 & 0.17     \\
         $M_Z$ [GeV]                    & 91.1876& 0.0021   \\
    \hline
    \end{tabular}
    \caption{\it Values of key input parameters used in our EW fits. For $M_Z$, $m_H$ and $m_{top}$, PDG values are adopted~\cite{Zyla:2020zbs}. Note, however, the inflated uncertainty for $m_{\rm top}$: it reflects a conservative choice along the lines of what is well-motivated, e.g., in Ref.~\cite{deBlas:2021wap}. Values of $\alpha_s(M_Z)$ and $\Delta\alpha^{(5)}_{had}(M_Z)$, as well as some minor inputs controlling intrinsic theory uncertainties, are gathered from Ref.~\cite{deBlas:2021wap}. Values of $G_F=1.1663787\times10^{-5}$ and $\alpha(0)=1/137.035999139$ are used  as fixed parameters given the precision to which they are measured~\cite{Zyla:2020zbs}.}
    \label{tab:params}
\end{table}

We explore several scenarios in this work, each for the two $M_W$ averages quoted in \autoref{eq:mw_values} separately:
\begin{itemize}
\itemsep0em
    \item {\bf SM}: We perform an SM fit of the EWPO measurements and report the corresponding $IC$ in \autoref{tab:IC}. We also predict the SM expectation of $M_W=80.355\pm0.0008$ GeV that is shown with a black band in \autoref{fig:fig1}. 
    \item {\bf 1 parameter fits:} We isolate each of the four Wilson coefficients, $C_{HWB}$, $C_{HD}$, $C_{LL}$ and $C_{HL}^{(3)}$ and study the constraints of the new measurement on each of them separately. 
    \item {\bf 2 parameter fits:} We perform two-parameter fits with $C_{HWB}$ and $C_{HD}$ to study the effects of solely oblique contributions reflected through the $S$ and $T$ parameters. 
    \item {\bf 4 parameter fits:} We then take a look at four-parameter fits of $C_{HWB}$, $C_{HD}$, $C_{LL}$ and $C_{HL}^{(3)}$ to gauge the constraints set by the new measurement on these Wilson coefficients when they are present simultaneously.
    \item {\bf \boldmath $\Delta\alpha^{(5)}_{had}(M_Z)$ study:} In the spirit of what was originally done in Ref.~\cite{Passera:2008jk}, to investigate any possible interesting effects in relation to the long-standing anomaly $(g-2)_\mu$, we perform fits where we exclude the current measurements of $\Delta\alpha^{(5)}_{had}(M_Z)$ with an aim to compare the SM prediction for the hadron vacuum polarization contribution to $\alpha$ against the one predicted in the best-fit NP scenario found.
\end{itemize}

\section{Lessons from the SMEFT}
\label{sec:lessons}

\begin{figure}[t]
  \centering
  \includegraphics[width=0.5\textwidth]{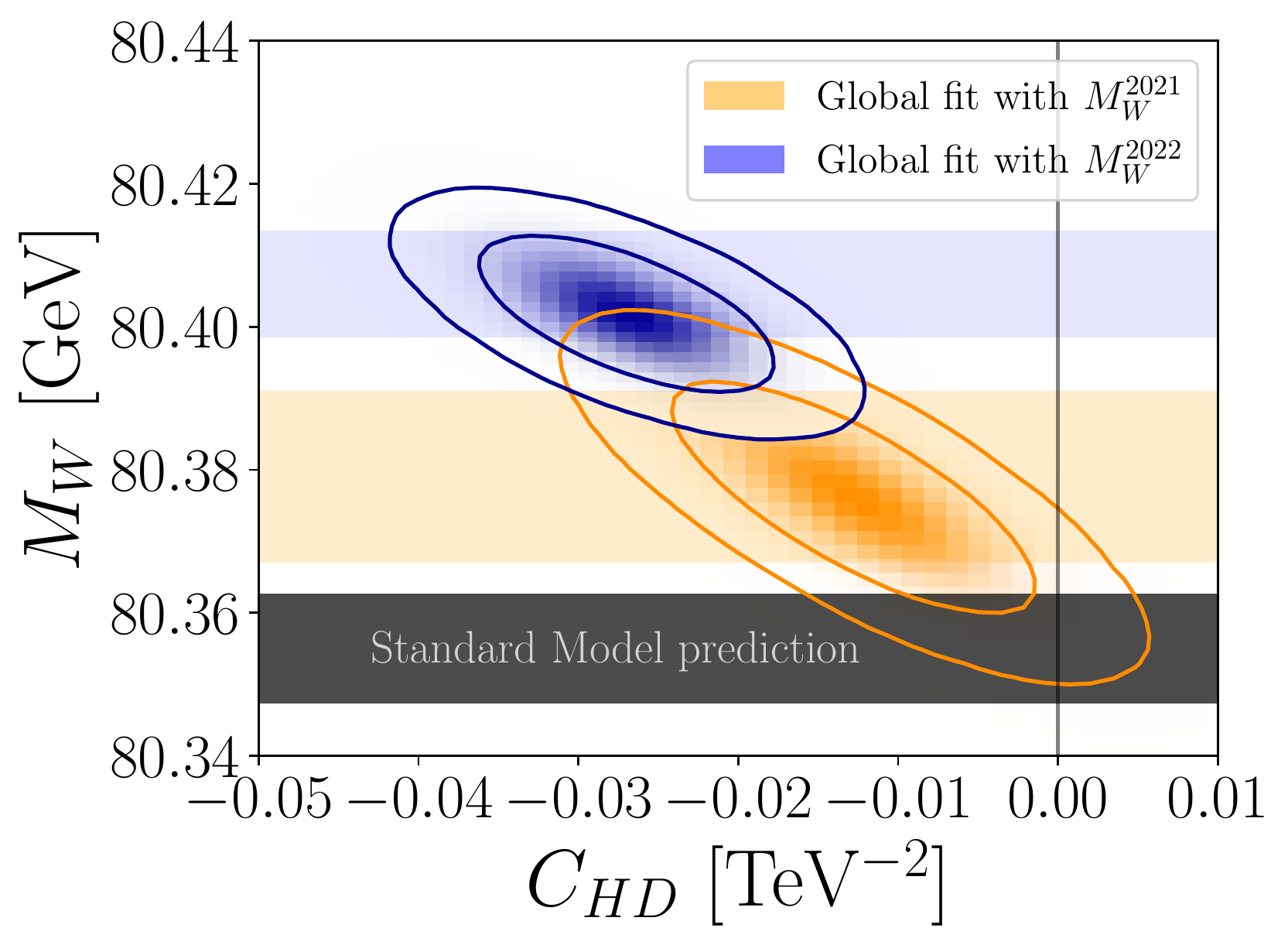}
  \caption{\it A ``before and after'' plot showing the constraints from EWPO in the $M_W$ vs. $C_{HD}$ plane. The orange and blue horizontal bands correspond to the global fits for $M_W$ without the recent CDF II measurement (assuming the absence of NP contributions) and the current global fit including the recent measurement, respectively. The 68\% (shaded region) and 95\% probability contours highlight the NP contribution to the EWPO from $\mathcal{O}_{HD}$ only. Remarkably, the new CDF II measurement yields a deviation of this NP Wilson coefficient from 0 at 4.5$\sigma$. The black band is the SM prediction obtained by using all other EWPO data in the fit except $M_W$.}
  \label{fig:fig1}
\end{figure}

\begin{figure}[t]
  \centering
  \includegraphics[width=0.5\textwidth]{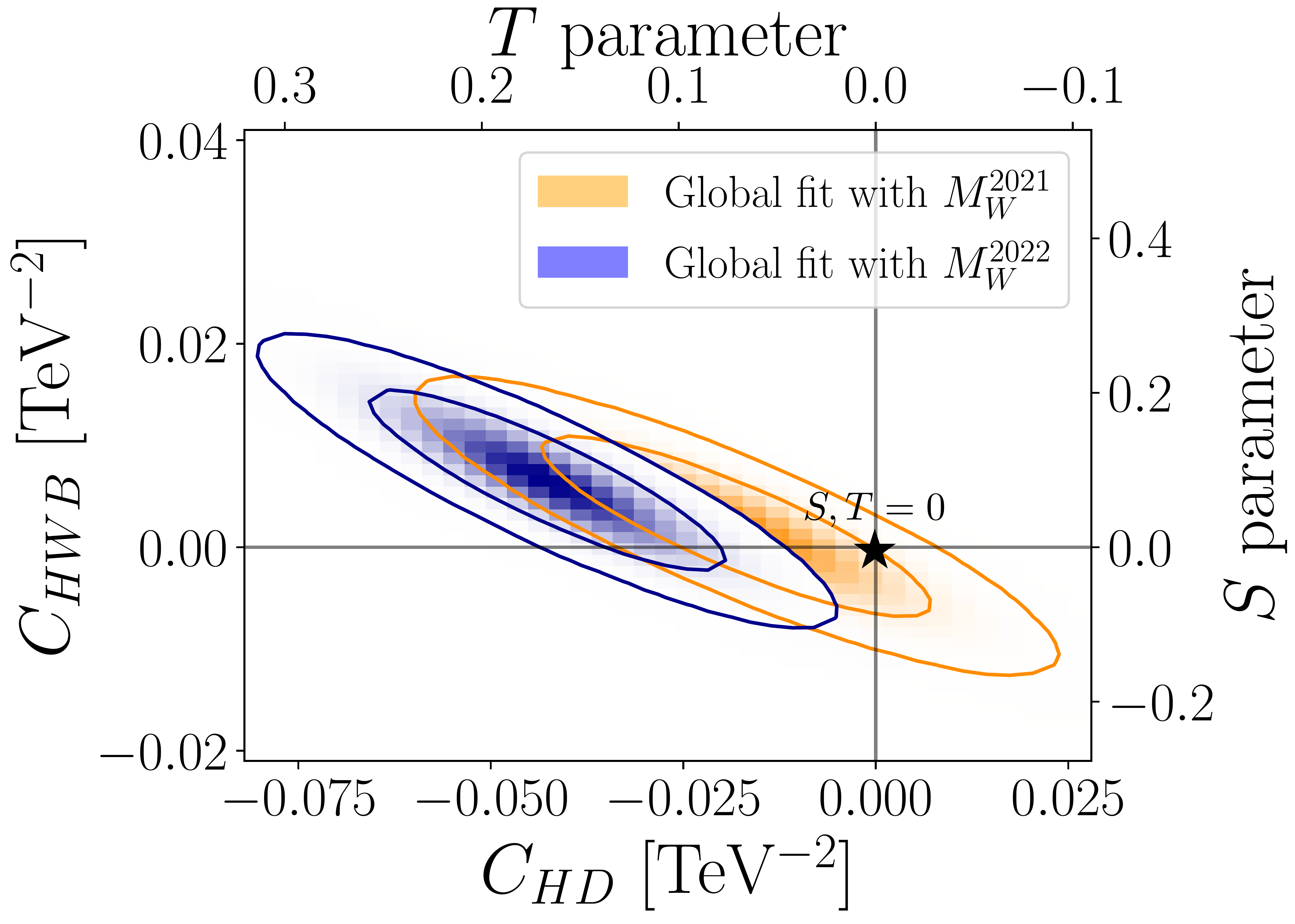}
  \caption{\it The results of a two-parameter fit with the Wilson coefficients $C_{HWB}$ and $C_{HD}$ allowing for the encapsulation of NP effects that are purely oblique, namely captured by the Peskin-Takeuchi parameters $S$ and $T$~\cite{Peskin:1991sw} which are linearly dependent on $C_{HWB}$ and $C_{HD}$ at dimension six in the SMEFT. The 68\% (shaded region) and 95\% probability contours in orange and blue correspond to fits excluding and including the recent CDF II measurement of the $M_W$ mass, respectively. A clear hint of custodial symmetry breaking can be seen from this plot with $T\ne 0$ at more than $3\sigma$ from the SM limit, marked with the dark star.}
  \label{fig:fig2}
\end{figure}

\begin{table}[t]
    \centering
    {\footnotesize
    \begin{tabular}{c|r@{ $\pm$ }l|c|r@{ $\pm$ }l|c}
    \toprule
    \multirow{2}{*}{scenario} &
    \multicolumn{2}{c|}{mean$\pm$ error} &
    \multirow{2}{*}{$IC^{2021}$} &
    \multicolumn{2}{c|}{mean$\pm$ error} &
    \multirow{2}{*}{$IC^{2022}$} \\
    &  \multicolumn{2}{c|}{2021} && \multicolumn{2}{c|}{2022} &\\
    \hline
    SM              & \multicolumn{2}{c|}{--}           & 22   & \multicolumn{2}{c|}{--} & 52 \\
    \hline
    \multicolumn{7}{c}{single Wilson coefficient fits}\\
    \hline
    $C_{HWB}$       &(-3.6&2.6)$\times 10^{-3}$ & 21    &(-8.5&2.3)$\times 10^{-3}$ & 33    \\
    $C_{HD}$        &(-1.3&0.7)$\times 10^{-2}$ & 20    &\color{BrickRed}{(-2.7}&\color{BrickRed}{0.6)$\times 10^{-2}$} & \color{BrickRed}{23}    \\
    $C_{LL}$        &(6.6&6.1)$\times 10^{-3}$  & 22    &(1.5&0.6)$\times 10^{-2}$  & 46    \\
    $C_{HL}^{(3)}$  &(-5.8&3.6)$\times 10^{-3}$ & 20    &(1.2&0.3)$\times 10^{-2}$  & 37    \\
    \hline
    \multicolumn{7}{c}{two Wilson coefficient fit}\\
    \hline
    $C_{HWB}$       &(2.1&5.9)$\times 10^{-3}$  &\multirow{2}{*}{22}&(6.6&5.8)$\times 10^{-3}$&\multirow{2}{*}{24}\\
    $C_{HD}$        &(-1.8&1.7)$\times 10^{-2}$ &       &(-4.2&1.5)$\times 10^{-2}$ &       \\
    \hline
    \multicolumn{7}{c}{four Wilson coefficient fit}\\
    \hline
    $C_{HWB}$       &(1.4&7.5)$\times 10^{-3}$&\multirow{4}{*}{27}&(1.4&7.6)$\times 10^{-3}$&\multirow{4}{*}{27}\\
    $C_{HD}$        &(-1.6&1.7)$\times 10^{-2}$ &       &(-3.8&1.5)$\times 10^{-2}$ &       \\ 
    $C_{LL}$        &(-1.0&1.4)$\times 10^{-2}$ &       &(-2.1&1.4)$\times 10^{-2}$ &       \\
    $C_{HL}^{(3)}$  &(-6.4&7.3)$\times 10^{-2}$ &       &(-6.4&7.3)$\times 10^{-3}$ &       \\
    \hline
    \end{tabular}
    }
    \caption{\it The $IC$s of the various fit scenarios. A lower number indicates a better fit to data. The details of the implications of these numbers are discussed in the text. All values for the Wilson coefficients are expressed in TeV$^{-2}$. A significance of NP greater than 4.5$\sigma$ is seen in the single Wilson coefficient fit with $C_{HD}$.}
    \label{tab:IC}
\end{table}

\begin{figure*}[t]
  \centering
  \includegraphics[width=0.95\textwidth]{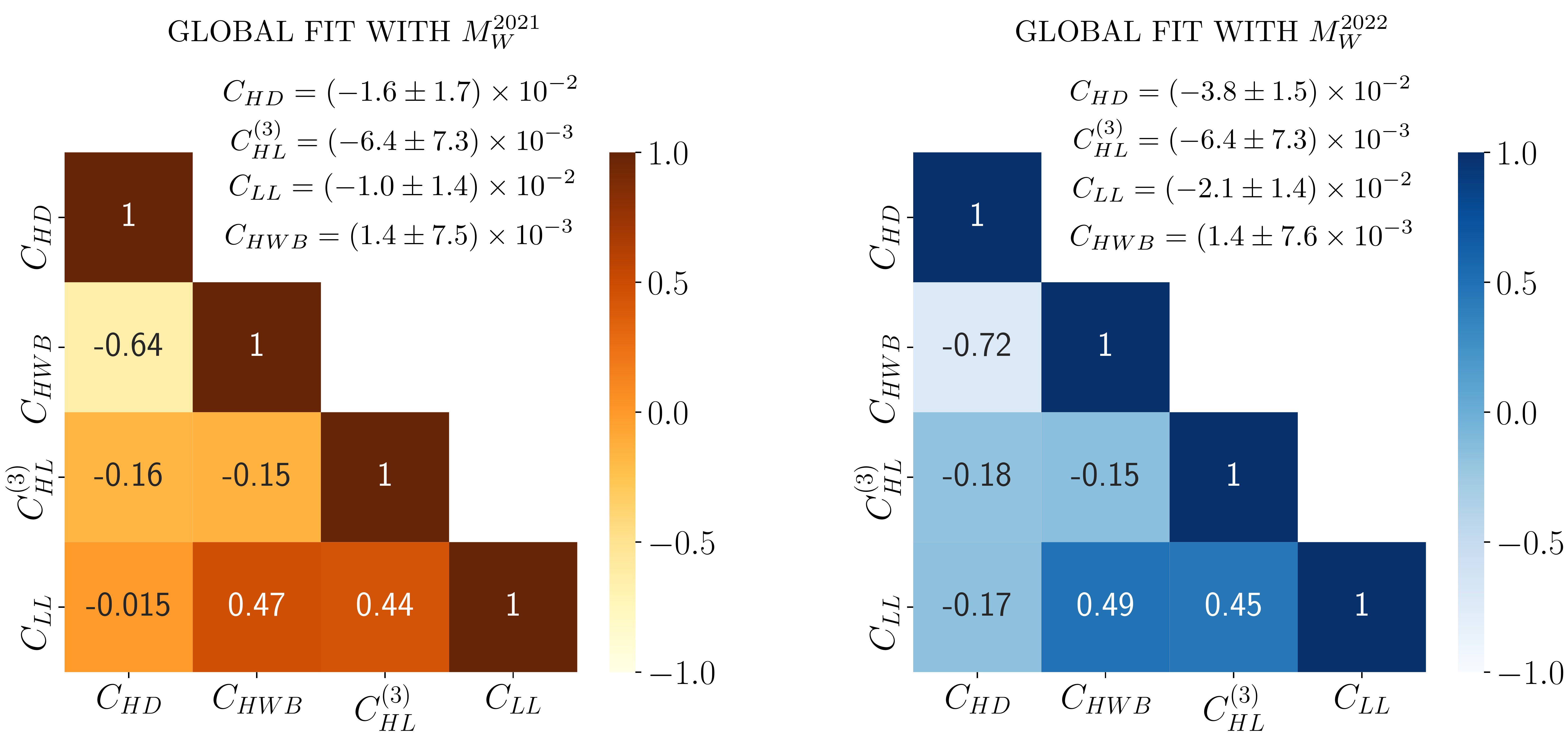}
  \caption{\it Outcome of the four-parameter fit using $C_{HWB}$, $C_{HD}$, $C_{LL}$ and $C_{HL}^{(3)}$ showing the mean and error of each coefficient from the fit along with the correlation matrix. The panel on the left corresponds to the global fit using the $M_W$ value before the recent CDF II measurement and the panel on the right represents the global fit including the recent CDF II measurement. The Wilson coefficient most affected by the recent measurement is $C_{HD}$, which even in this multi-dimensional NP case remains non-zero at the 2.5$\sigma$ level, with a higher central value and reduced error than before.}
  \label{fig:fig3}
\end{figure*}

We report the $IC$ for several cases in \autoref{tab:IC} from which we get a clear picture of what the effects of the new CDF II measurement are on the SM and NP scenarios.\vspace{0.2cm}

\noindent{\bf SM}: The recent measurement of $M_W$ has increased the tension between the data and the SM prediction. $IC^{2022}_{\rm SM}$ is much higher than $IC^{2021}_{\rm SM}$ showcasing this increase in discrepancy between the data and the model.\vspace{0.1cm}

\noindent{\bf 2021}: If we focus on $IC^{2021}_{NP}$, the addition of single NP Wilson coefficients did not improve the fit (barring small differences that are not significant) over an SM fit with $M_W^{2021}$. The same can be said for the two-parameter fit where we see no improvement over the SM fit. The four-parameter fit with all four Wilson coefficients is actually worse than any of the fits because the fit does not improve and it gets penalized for having a larger number of parameters. All in all, the fits with $M_W^{2021}$ do not show any hints of NP contributions within the framework we consider.\vspace{0.1cm}

\noindent{\bf 2022}: Now taking a look at the $IC^{2022}_{NP}$ we see a very different pattern. The fit with only $C_{HD}$ performs much better than any other fits including NP contributions. In fact, it alleviates almost all the tension that has been generated by the new CDF II measurement of the $W$ mass yielding an $IC$ very close that $IC^{2021}_{\rm SM}$. The other single-parameter fits perform much worse than they do with $M_W^{2021}$. The $IC^{2022}_{NP}$ of the two-parameter fit with $C_{HWB}$ and $C_{HD}$ performs slightly worse than the single parameter $C_{HD}$ fit, presumably because of no improvement in the goodness of fit while being penalized for the increase in the number of parameters. The $IC^{2022}_{NP}$ of the four-parameter fit is the same as the $IC^{2021}_{NP}$ for the same fit showing no marked improvement. \vspace{0.2cm}

The most important message that we wish to highlight in this work is the fact that the new CDF II measurement of $W$ mass distinctly points at a hint of NP and preferably from a UV model that generates $C_{HD}$ with a significance of greater than 4.5$\sigma$. The presence of the other Wilson coefficients does not significantly worsen the fit and an NP model that generates those can also be accommodated by the current EWPO data.

In \autoref{fig:fig1} we show the constraints on $C_{HD}$ and their correlation with the $W$-mass measurements. The orange band and curves correspond to global fits using $M_W^{2021}$ in the SM and assuming the presence of NP manifested through $C_{HD}$, respectively. The blue band and curves are the same for the global fit using  $M_W^{2022}$. While, there are no significant hint of NP contribution when using  $M_W^{2021}$, the significance of the same increases drastically with $M_W^{2022}$. The SM prediction is marked in dark.

In \autoref{fig:fig2} we show the results of the two-parameter fit to motivate a discussion of hints of custodial symmetry breaking, represented by a non-zero $T$ parameter. The orange curves show the fit including  $M_W^{2021}$ and the blue ones the fit including  $M_W^{2022}$. In the former case the point $S=T=0$ is included in the 1$\sigma$ (shaded) region while in the latter case it is significantly outside the 2$\sigma$ region. The $S$ and $T$ values in the scenario with $M_W^{2021}$:
\begin{equation*}
    S = 0.027\pm0.076 \, , \ T = 0.070\pm0.066 \ \ (\rho = 0.898) \, , \\
\end{equation*}
can be compared with the new ones adopting $M_W^{2022}$:
\begin{equation*}
    S  =  0.086\pm0.076 \, , \ T = 0.167\pm0.059 \ \  (\rho = 0.916) \, ,
\end{equation*}
with the latter set showing significant hints for sources of custodial-symmetry violation beyond the SM.

Finally, in \autoref{fig:fig3} we show the results from varying all four Wilson coefficients simultaneously in the fit. The left panel is the global fit including $M_W^{2021}$ and the right panel is the global fit including $M_W^{2022}$. The current measurement increases the correlation $\rho$ between $C_{HWB}$ and $C_{HD}$, and between $C_{HD}$, and $C_{LL}$, reshuffling the significance for NP also in the latter, but leaving the general structure of the correlation pattern almost unchanged.

\section{Discussion.} 
\label{sec:discussion}

The recent $W$-mass measurement by CDF can have far-reaching consequences on our understanding of the dynamics that govern our Universe and could pave a path to the discovery of possible BSM physics. Here we would like to conclude by giving a brief overview of further implications that this measurement can actually have.

\subsection{Notes on implications for $(g-2)_{\mu}$}
The recent measurement of the muon anomalous magnetic moment at the Fermilab~\cite{Muong-2:2021ojo} adds to the discrepancy that already existed from an earlier BNL E821 measurement~\cite{Muong-2:2006rrc} bringing up the tension with the SM estimate to about 4.2$\sigma$~\cite{Muong-2:2021ojo,Aoyama:2020ynm}. It should be noted that the recent lattice results from the BMW collaboration~\cite{Borsanyi:2020mff} have not been included in the world average and considering them as a realistic contribution~\cite{Davier:2010nc,Davier:2017zfy,Davier:2019can} to the HVP of the photon reduces the tension with the SM estimate of $(g-2)_\mu$. On a different note, the same $(g-2)_e$ may represent a puzzle within the SM, given the opposite sign observed with respect to the muon counterpart~\cite{Keshavarzi:2020bfy}.

The computation of $a_\mu\equiv(g-2)_\mu/2$, is quite sensitive to the estimated value of $\Delta\alpha^{(5)}_{had}(M_Z)$ where an increase in the value of $\Delta\alpha^{(5)}_{had}(M_Z)$ may reduce the discrepancy of $a_\mu$ from its experimental measurement if naively translated into a rescaling of the low-energy $e^+ e^-\to $ hadrons cross-section. Given the new measurement from CDF II some exploration in this direction is warranted. 

\begin{figure}[t!]
  \centering
  \includegraphics[width=0.49\textwidth]{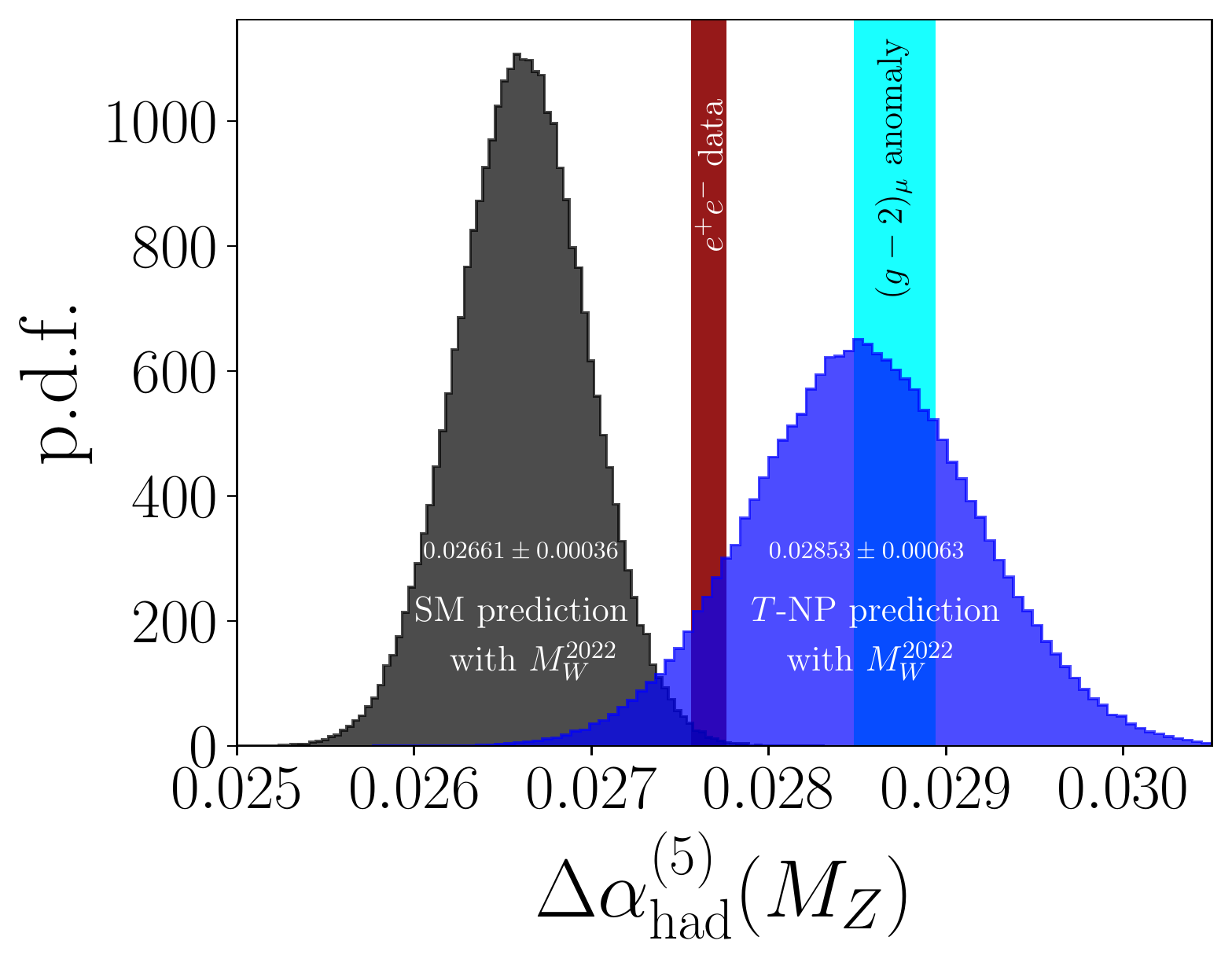}
  \caption{\it Given its intimate relation with $(g-2)_\mu$~\cite{Keshavarzi:2020bfy}, we show how $\Delta\alpha^{(5)}_{had}(M_Z)$ is affected by the new measurement of $M_W$. A clear tension lies in the measurement of $\Delta\alpha^{(5)}_{had}(M_Z)$ from $e^+e^-$ data represented by the dark red band and the value expected from the measurement of $(g-2)_{\mu}$ as estimated in Ref.~\cite{Crivellin:2020zul} and represented here by the light blue band. The black distribution represents the SM prediction using all current measurements of EWPO except the $\Delta\alpha^{(5)}_{had}(M_Z)$ measurement which shows an increased tension. The dark blue distribution corresponds to the same in case of NP from $C_{HD}$ where one can see that the tension is resolved.}
  \label{fig:fig4}
\end{figure}

In \autoref{fig:fig4} we see that there is a strong tension between the measurement of $\Delta\alpha^{(5)}_{had}(M_Z)$ from $e^+e^-$ data and the estimate from $(g-2)_\mu$ measurement. If we assume the absence of NP contributions and perform an SM fit with $M_W^{2022}$ we find that the discrepancy gets even worse with the mean of $\Delta\alpha^{(5)}_{had}(M_Z)$ getting even lower. However, if we allow for the presence of NP through a non-zero $C_{HD}$ the discrepancy between the value of $\Delta\alpha^{(5)}_{had}(M_Z)$ estimated from the global fit and that from the $(g-2)_\mu$ measurement can be completely alleviated. It is to be noted that to arrive at the estimate of $\Delta\alpha^{(5)}_{had}(M_Z)$ from the EWPO data we did not use the measured value of $\Delta\alpha^{(5)}_{had}(M_Z)$ from the $e^+e^-$ data but rather determined it indirectly from the combination of all other measurements. We have labeled in \autoref{fig:fig4} such a prediction as $T$-NP prediction, given the underlying custodial-symmetry violation implied in the scenario.

\subsection{Notes on implications for $B$ anomalies}
Hints for Lepton Universality Violation (LUV), stemming in particular from the study of $B$-meson decays via flavour-changing neutral-current processes, are close to the level of 5$\sigma$ significance in favor of BSM physics~\cite{Isidori:2021vtc,Ciuchini:2021smi}. A joint broad analysis of EWPO, LUV observables $R_K$ and $R_K^*$, angular observables in $b\to s \ell^+\ell^-$ decays and $B_{s,d}\to \ell^+\ell^-$ was initiated in Ref.~\cite{Alasfar:2020mne}. There, it was shown that in a flavor non-universal scenario, the constraints from EWPO and the hints for NP from Flavor Physics can be simultaneously accommodated within a new $Z^{\prime}$ gauge boson at the scale of $\Lambda_{NP}$ of a few TeV. It is known that models with a new $Z^\prime$ can generate custodial symmetry breaking~\cite{Cacciapaglia:2006pk} opening a path to tying together the flavour anomalies with the hint of custodial symmetry breaking. Note that the inference of $C_{HD}$ highlighted in red in \autoref{tab:IC} underlies naively an NP scale of about 6~TeV assuming tree-level new dynamics with $\mathcal{O}(1)$ couplings. It would be, therefore, interesting to combine the hint for LUV with the violation of custodial symmetry inferred in this work. A systematic study that would take into account also effects from the SM renormalization group at the one-loop level would be warranted in the spirit of what was carried out in Ref.~\cite{Alasfar:2020mne} and is left for future work.

\subsection{Notes on model realizations}
The preservation of custodial symmetry has been a guiding principle for the construction of many realistic models of NP. Hence, the hint of breaking of custodial symmetry will have strong implications on BSM model building. We would like to leave a comment on the implication of the new world average of the $W$-boson mass on a general 2HDM model. Extensive studies of contributions to custodial symmetry breaking in the 2HDM have been conducted over the years~\cite{Pomarol:1993mu,Gerard:2007kn,Haber:2010bw,Funk:2011ad}. The imposition of custodial symmetry implies that the mass of the charged Higgs should be degenerate with the mass of the pseudoscalar Higgs, i.e., $m_{H^\pm}=m_{A^0}$ in the extended Higgs sector when imposing the symmetry through an additional constraint on the CP-conserving scalar potential. It is also possible to impose custodial symmetry by asserting $m_{H^\pm}=m_{H}$ or $m_{H^\pm}=m_{H^0}$ where $H$ is the light Higgs identified with the 125 GeV Higgs and $H^0$ is the heavy Higgs scalar~\cite{Gerard:2007kn,Haber:2010bw}.  The former condition is already ruled out due to constraints on the charged Higgs from $b\to s\gamma$ inclusive decay.

An observation of custodial-symmetry breaking would imply that the Higgs mass spectrum in a 2HDM will have bounds on possible degeneracies in the physical states of the scalar and pseudoscalar bosons requiring a splitting between these states depending on the model construction. This numerically and conceptually changes the primary impetus of imposing custodial symmetry on the 2HDM models and moves towards a necessity to generate the breaking of the symmetry. 

In general, models with additional scalar multiplets of $SU(2)_L$ have the potential to generate $O_{HD}$ which will contribute to the breaking of custodial symmetry as can be seen from \autoref{eq:ST}. In Ref.~\cite{Dawson:2017vgm} it was shown that while real singlets with explicit or spontaneous $Z_2$ breaking cannot generate $O_{HD}$, real and complex triplets and quartets with hypercharge 3/2 or 1/2 and $Z_2$ symmetry breaking can generate $O_{HD}$ leading to possible new sources of custodial-symmetry violation. A more detailed investigation of these models is desirable but outside the scope of this work.

\subsection{Final Notes}
The measurement of the $W$-boson mass is notoriously difficult and experimentally challenging, requiring a perfect understanding of detector properties and simulations, as well as a non-trivial cross-talk among theory and experimental communities. While we look forward to new updates on this matter, as well as decisive steps in the future experimental EW precision program, we point out that at present a naive global combination of $W$-boson mass, namely $M_W = 80.4060 \pm 0.0075$ GeV, provides robust support, at up to 4.5$\sigma$, for tree-level (relatively strongly coupled) new dynamics at the scale of few TeV, entering in the Peskin-Takeuchi $T$ parameter, signaling a new source of violation of custodial symmetry that we can hope to discover in the near future. The most viable physics scenario in the SMEFT framework that we consider also leaves open the possibility of alleviating another discrepancy that surfaced in the recent past in the lepton sector, namely, the measurement of the anomalous magnetic moment of the muon. The preferred NP scenario also holds potential implications for the $B$-physics anomalies that have driven a lot of recent work and theoretical developments in the recent past. Hence, we see a true opportunity to tie together different sectors of the standard theory in order to learn more about its UV completion and carry out pivotal progress in the fundamental understanding of Nature.

~

\begin{acknowledgments}
The authors are grateful to L. Silvestrini for all the details provided regarding the analysis in Ref.~\cite{deBlas:2021wap} and J. de Blas for his comments and helpful insights. M.V. acknowledges P. Meade for stimulating discussion and A. Florio for a very needed coffee. Upon completion of this work, a few related studies appeared~\cite{Fan:2022dck,Lu:2022bgw,Athron:2022qpo,Strumia:2022qkt,blas2022impact}, some of which highlight similar conclusions as ours. The work of A.P. is funded by Volkswagen Foundation within the initiative ``Corona Crisis and Beyond -- Perspectives for Science, Scholarship and Society'', grant number 99091. The work of M.V. is supported by the Simons Foundation under the Simons Bridge for Postdoctoral Fellowships at SCGP and YITP, award number 815892.
\end{acknowledgments}

\bibliography{apssamp}

\begin{thebibliography}{83}%
\makeatletter
\providecommand \@ifxundefined [1]{%
 \@ifx{#1\undefined}
}%
\providecommand \@ifnum [1]{%
 \ifnum #1\expandafter \@firstoftwo
 \else \expandafter \@secondoftwo
 \fi
}%
\providecommand \@ifx [1]{%
 \ifx #1\expandafter \@firstoftwo
 \else \expandafter \@secondoftwo
 \fi
}%
\providecommand \natexlab [1]{#1}%
\providecommand \enquote  [1]{``#1''}%
\providecommand \bibnamefont  [1]{#1}%
\providecommand \bibfnamefont [1]{#1}%
\providecommand \citenamefont [1]{#1}%
\providecommand \href@noop [0]{\@secondoftwo}%
\providecommand \href [0]{\begingroup \@sanitize@url \@href}%
\providecommand \@href[1]{\@@startlink{#1}\@@href}%
\providecommand \@@href[1]{\endgroup#1\@@endlink}%
\providecommand \@sanitize@url [0]{\catcode `\\12\catcode `\$12\catcode
  `\&12\catcode `\#12\catcode `\^12\catcode `\_12\catcode `\%12\relax}%
\providecommand \@@startlink[1]{}%
\providecommand \@@endlink[0]{}%
\providecommand \url  [0]{\begingroup\@sanitize@url \@url }%
\providecommand \@url [1]{\endgroup\@href {#1}{\urlprefix }}%
\providecommand \urlprefix  [0]{URL }%
\providecommand \Eprint [0]{\href }%
\providecommand \doibase [0]{https://doi.org/}%
\providecommand \selectlanguage [0]{\@gobble}%
\providecommand \bibinfo  [0]{\@secondoftwo}%
\providecommand \bibfield  [0]{\@secondoftwo}%
\providecommand \translation [1]{[#1]}%
\providecommand \BibitemOpen [0]{}%
\providecommand \bibitemStop [0]{}%
\providecommand \bibitemNoStop [0]{.\EOS\space}%
\providecommand \EOS [0]{\spacefactor3000\relax}%
\providecommand \BibitemShut  [1]{\csname bibitem#1\endcsname}%
\let\auto@bib@innerbib\@empty
\bibitem [{\citenamefont {Aaltonen}\ \emph {et~al.}(2022)\citenamefont
  {Aaltonen} \emph {et~al.}}]{CDF:2022hxs}%
  \BibitemOpen
  \bibfield  {author} {\bibinfo {author} {\bibfnamefont {T.}~\bibnamefont
  {Aaltonen}} \emph {et~al.} (\bibinfo {collaboration} {CDF}),\ }\bibfield
  {title} {\bibinfo {title} {{High-precision measurement of the $W$ boson mass
  with the CDF II detector}},\ }\href {https://doi.org/10.1126/science.abk1781}
  {\bibfield  {journal} {\bibinfo  {journal} {Science}\ }\textbf {\bibinfo
  {volume} {376}},\ \bibinfo {pages} {170} (\bibinfo {year}
  {2022})}\BibitemShut {NoStop}%
\bibitem [{\citenamefont {Aaltonen}\ \emph {et~al.}(2012)\citenamefont
  {Aaltonen} \emph {et~al.}}]{CDF:2012gpf}%
  \BibitemOpen
  \bibfield  {author} {\bibinfo {author} {\bibfnamefont {T.}~\bibnamefont
  {Aaltonen}} \emph {et~al.} (\bibinfo {collaboration} {CDF}),\ }\bibfield
  {title} {\bibinfo {title} {{Precise measurement of the $W$-boson mass with
  the CDF II detector}},\ }\href
  {https://doi.org/10.1103/PhysRevLett.108.151803} {\bibfield  {journal}
  {\bibinfo  {journal} {Phys. Rev. Lett.}\ }\textbf {\bibinfo {volume} {108}},\
  \bibinfo {pages} {151803} (\bibinfo {year} {2012})},\ \Eprint
  {https://arxiv.org/abs/1203.0275} {arXiv:1203.0275 [hep-ex]} \BibitemShut
  {NoStop}%
\bibitem [{\citenamefont {Aaboud}\ \emph {et~al.}(2018)\citenamefont {Aaboud}
  \emph {et~al.}}]{ATLAS:2017rzl}%
  \BibitemOpen
  \bibfield  {author} {\bibinfo {author} {\bibfnamefont {M.}~\bibnamefont
  {Aaboud}} \emph {et~al.} (\bibinfo {collaboration} {ATLAS}),\ }\bibfield
  {title} {\bibinfo {title} {{Measurement of the $W$-boson mass in pp
  collisions at $\sqrt{s}=7$ TeV with the ATLAS detector}},\ }\href
  {https://doi.org/10.1140/epjc/s10052-017-5475-4} {\bibfield  {journal}
  {\bibinfo  {journal} {Eur. Phys. J. C}\ }\textbf {\bibinfo {volume} {78}},\
  \bibinfo {pages} {110} (\bibinfo {year} {2018})},\ \bibinfo {note} {[Erratum:
  Eur.Phys.J.C 78, 898 (2018)]},\ \Eprint {https://arxiv.org/abs/1701.07240}
  {arXiv:1701.07240 [hep-ex]} \BibitemShut {NoStop}%
\bibitem [{\citenamefont {Abazov}\ \emph {et~al.}(2014)\citenamefont {Abazov}
  \emph {et~al.}}]{D0:2013jba}%
  \BibitemOpen
  \bibfield  {author} {\bibinfo {author} {\bibfnamefont {V.~M.}\ \bibnamefont
  {Abazov}} \emph {et~al.} (\bibinfo {collaboration} {D0}),\ }\bibfield
  {title} {\bibinfo {title} {{Measurement of the $W$ boson mass with the D0
  detector}},\ }\href {https://doi.org/10.1103/PhysRevD.89.012005} {\bibfield
  {journal} {\bibinfo  {journal} {Phys. Rev. D}\ }\textbf {\bibinfo {volume}
  {89}},\ \bibinfo {pages} {012005} (\bibinfo {year} {2014})},\ \Eprint
  {https://arxiv.org/abs/1310.8628} {arXiv:1310.8628 [hep-ex]} \BibitemShut
  {NoStop}%
\bibitem [{\citenamefont {Aaij}\ \emph {et~al.}(2022)\citenamefont {Aaij} \emph
  {et~al.}}]{LHCb:2021bjt}%
  \BibitemOpen
  \bibfield  {author} {\bibinfo {author} {\bibfnamefont {R.}~\bibnamefont
  {Aaij}} \emph {et~al.} (\bibinfo {collaboration} {LHCb}),\ }\bibfield
  {title} {\bibinfo {title} {{Measurement of the $W$ boson mass}},\ }\href
  {https://doi.org/10.1007/JHEP01(2022)036} {\bibfield  {journal} {\bibinfo
  {journal} {JHEP}\ }\textbf {\bibinfo {volume} {01}},\ \bibinfo {pages}
  {036}},\ \Eprint {https://arxiv.org/abs/2109.01113} {arXiv:2109.01113
  [hep-ex]} \BibitemShut {NoStop}%
\bibitem [{\citenamefont {Zyla}\ \emph {et~al.}(2020)\citenamefont {Zyla} \emph
  {et~al.}}]{Zyla:2020zbs}%
  \BibitemOpen
  \bibfield  {author} {\bibinfo {author} {\bibfnamefont {P.}~\bibnamefont
  {Zyla}} \emph {et~al.} (\bibinfo {collaboration} {Particle Data Group}),\
  }\bibfield  {title} {\bibinfo {title} {{Review of Particle Physics}},\ }\href
  {https://doi.org/10.1093/ptep/ptaa104} {\bibfield  {journal} {\bibinfo
  {journal} {PTEP}\ }\textbf {\bibinfo {volume} {2020}},\ \bibinfo {pages}
  {083C01} (\bibinfo {year} {2020})},\ \bibinfo {note} {and 2021
  update}\BibitemShut {NoStop}%
\bibitem [{\citenamefont {Sirlin}(1980)}]{Sirlin:1980nh}%
  \BibitemOpen
  \bibfield  {author} {\bibinfo {author} {\bibfnamefont {A.}~\bibnamefont
  {Sirlin}},\ }\bibfield  {title} {\bibinfo {title} {{Radiative Corrections in
  the $SU(2)_L \times U(1)$ Theory: A Simple Renormalization Framework}},\
  }\href {https://doi.org/10.1103/PhysRevD.22.971} {\bibfield  {journal}
  {\bibinfo  {journal} {Phys. Rev. D}\ }\textbf {\bibinfo {volume} {22}},\
  \bibinfo {pages} {971} (\bibinfo {year} {1980})}\BibitemShut {NoStop}%
\bibitem [{\citenamefont {Marciano}\ and\ \citenamefont
  {Sirlin}(1980)}]{Marciano:1980pb}%
  \BibitemOpen
  \bibfield  {author} {\bibinfo {author} {\bibfnamefont {W.~J.}\ \bibnamefont
  {Marciano}}\ and\ \bibinfo {author} {\bibfnamefont {A.}~\bibnamefont
  {Sirlin}},\ }\bibfield  {title} {\bibinfo {title} {{Radiative Corrections to
  Neutrino Induced Neutral Current Phenomena in the $SU(2)_L \times U(1)$
  Theory}},\ }\href {https://doi.org/10.1103/PhysRevD.22.2695} {\bibfield
  {journal} {\bibinfo  {journal} {Phys. Rev. D}\ }\textbf {\bibinfo {volume}
  {22}},\ \bibinfo {pages} {2695} (\bibinfo {year} {1980})},\ \bibinfo {note}
  {[Erratum: Phys.Rev.D 31, 213 (1985)]}\BibitemShut {NoStop}%
\bibitem [{\citenamefont {Djouadi}\ and\ \citenamefont
  {Verzegnassi}(1987)}]{Djouadi:1987gn}%
  \BibitemOpen
  \bibfield  {author} {\bibinfo {author} {\bibfnamefont {A.}~\bibnamefont
  {Djouadi}}\ and\ \bibinfo {author} {\bibfnamefont {C.}~\bibnamefont
  {Verzegnassi}},\ }\bibfield  {title} {\bibinfo {title} {{Virtual Very Heavy
  Top Effects in LEP / SLC Precision Measurements}},\ }\href
  {https://doi.org/10.1016/0370-2693(87)91206-8} {\bibfield  {journal}
  {\bibinfo  {journal} {Phys. Lett. B}\ }\textbf {\bibinfo {volume} {195}},\
  \bibinfo {pages} {265} (\bibinfo {year} {1987})}\BibitemShut {NoStop}%
\bibitem [{\citenamefont {Djouadi}(1988)}]{Djouadi:1987di}%
  \BibitemOpen
  \bibfield  {author} {\bibinfo {author} {\bibfnamefont {A.}~\bibnamefont
  {Djouadi}},\ }\bibfield  {title} {\bibinfo {title} {{$\mathcal{O}(\alpha
  \alpha_s)$ Vacuum Polarization Functions of the Standard Model Gauge
  Bosons}},\ }\href {https://doi.org/10.1007/BF02812964} {\bibfield  {journal}
  {\bibinfo  {journal} {Nuovo Cim. A}\ }\textbf {\bibinfo {volume} {100}},\
  \bibinfo {pages} {357} (\bibinfo {year} {1988})}\BibitemShut {NoStop}%
\bibitem [{\citenamefont {Kniehl}(1990)}]{Kniehl:1989yc}%
  \BibitemOpen
  \bibfield  {author} {\bibinfo {author} {\bibfnamefont {B.~A.}\ \bibnamefont
  {Kniehl}},\ }\bibfield  {title} {\bibinfo {title} {{Two Loop Corrections to
  the Vacuum Polarizations in Perturbative QCD}},\ }\href
  {https://doi.org/10.1016/0550-3213(90)90552-O} {\bibfield  {journal}
  {\bibinfo  {journal} {Nucl. Phys. B}\ }\textbf {\bibinfo {volume} {347}},\
  \bibinfo {pages} {86} (\bibinfo {year} {1990})}\BibitemShut {NoStop}%
\bibitem [{\citenamefont {Halzen}\ and\ \citenamefont
  {Kniehl}(1991)}]{Halzen:1990je}%
  \BibitemOpen
  \bibfield  {author} {\bibinfo {author} {\bibfnamefont {F.}~\bibnamefont
  {Halzen}}\ and\ \bibinfo {author} {\bibfnamefont {B.~A.}\ \bibnamefont
  {Kniehl}},\ }\bibfield  {title} {\bibinfo {title} {{$\Delta$ r beyond one
  loop}},\ }\href {https://doi.org/10.1016/0550-3213(91)90319-S} {\bibfield
  {journal} {\bibinfo  {journal} {Nucl. Phys. B}\ }\textbf {\bibinfo {volume}
  {353}},\ \bibinfo {pages} {567} (\bibinfo {year} {1991})}\BibitemShut
  {NoStop}%
\bibitem [{\citenamefont {Kniehl}\ and\ \citenamefont
  {Sirlin}(1992)}]{Kniehl:1991gu}%
  \BibitemOpen
  \bibfield  {author} {\bibinfo {author} {\bibfnamefont {B.~A.}\ \bibnamefont
  {Kniehl}}\ and\ \bibinfo {author} {\bibfnamefont {A.}~\bibnamefont
  {Sirlin}},\ }\bibfield  {title} {\bibinfo {title} {{Dispersion relations for
  vacuum polarization functions in electroweak physics}},\ }\href
  {https://doi.org/10.1016/0550-3213(92)90232-Z} {\bibfield  {journal}
  {\bibinfo  {journal} {Nucl. Phys. B}\ }\textbf {\bibinfo {volume} {371}},\
  \bibinfo {pages} {141} (\bibinfo {year} {1992})}\BibitemShut {NoStop}%
\bibitem [{\citenamefont {Kniehl}\ and\ \citenamefont
  {Sirlin}(1993)}]{Kniehl:1992dx}%
  \BibitemOpen
  \bibfield  {author} {\bibinfo {author} {\bibfnamefont {B.~A.}\ \bibnamefont
  {Kniehl}}\ and\ \bibinfo {author} {\bibfnamefont {A.}~\bibnamefont
  {Sirlin}},\ }\bibfield  {title} {\bibinfo {title} {{On the effect of the $t
  \bar{t}$ threshold on electroweak parameters}},\ }\href
  {https://doi.org/10.1103/PhysRevD.47.883} {\bibfield  {journal} {\bibinfo
  {journal} {Phys. Rev. D}\ }\textbf {\bibinfo {volume} {47}},\ \bibinfo
  {pages} {883} (\bibinfo {year} {1993})}\BibitemShut {NoStop}%
\bibitem [{\citenamefont {Barbieri}\ \emph {et~al.}(1992)\citenamefont
  {Barbieri}, \citenamefont {Beccaria}, \citenamefont {Ciafaloni},
  \citenamefont {Curci},\ and\ \citenamefont {Vicere}}]{Barbieri:1992nz}%
  \BibitemOpen
  \bibfield  {author} {\bibinfo {author} {\bibfnamefont {R.}~\bibnamefont
  {Barbieri}}, \bibinfo {author} {\bibfnamefont {M.}~\bibnamefont {Beccaria}},
  \bibinfo {author} {\bibfnamefont {P.}~\bibnamefont {Ciafaloni}}, \bibinfo
  {author} {\bibfnamefont {G.}~\bibnamefont {Curci}},\ and\ \bibinfo {author}
  {\bibfnamefont {A.}~\bibnamefont {Vicere}},\ }\bibfield  {title} {\bibinfo
  {title} {{Radiative correction effects of a very heavy top}},\ }\href
  {https://doi.org/10.1016/0370-2693(92)91960-H} {\bibfield  {journal}
  {\bibinfo  {journal} {Phys. Lett. B}\ }\textbf {\bibinfo {volume} {288}},\
  \bibinfo {pages} {95} (\bibinfo {year} {1992})},\ \bibinfo {note} {[Erratum:
  Phys.Lett.B 312, 511--511 (1993)]},\ \Eprint
  {https://arxiv.org/abs/hep-ph/9205238} {arXiv:hep-ph/9205238} \BibitemShut
  {NoStop}%
\bibitem [{\citenamefont {Barbieri}\ \emph {et~al.}(1993)\citenamefont
  {Barbieri}, \citenamefont {Beccaria}, \citenamefont {Ciafaloni},
  \citenamefont {Curci},\ and\ \citenamefont {Vicere}}]{Barbieri:1992dq}%
  \BibitemOpen
  \bibfield  {author} {\bibinfo {author} {\bibfnamefont {R.}~\bibnamefont
  {Barbieri}}, \bibinfo {author} {\bibfnamefont {M.}~\bibnamefont {Beccaria}},
  \bibinfo {author} {\bibfnamefont {P.}~\bibnamefont {Ciafaloni}}, \bibinfo
  {author} {\bibfnamefont {G.}~\bibnamefont {Curci}},\ and\ \bibinfo {author}
  {\bibfnamefont {A.}~\bibnamefont {Vicere}},\ }\bibfield  {title} {\bibinfo
  {title} {{Two loop heavy top effects in the Standard Model}},\ }\href
  {https://doi.org/10.1016/0550-3213(93)90448-X} {\bibfield  {journal}
  {\bibinfo  {journal} {Nucl. Phys. B}\ }\textbf {\bibinfo {volume} {409}},\
  \bibinfo {pages} {105} (\bibinfo {year} {1993})}\BibitemShut {NoStop}%
\bibitem [{\citenamefont {Djouadi}\ and\ \citenamefont
  {Gambino}(1994)}]{Djouadi:1993ss}%
  \BibitemOpen
  \bibfield  {author} {\bibinfo {author} {\bibfnamefont {A.}~\bibnamefont
  {Djouadi}}\ and\ \bibinfo {author} {\bibfnamefont {P.}~\bibnamefont
  {Gambino}},\ }\bibfield  {title} {\bibinfo {title} {{Electroweak gauge bosons
  selfenergies: Complete QCD corrections}},\ }\href
  {https://doi.org/10.1103/PhysRevD.49.3499} {\bibfield  {journal} {\bibinfo
  {journal} {Phys. Rev. D}\ }\textbf {\bibinfo {volume} {49}},\ \bibinfo
  {pages} {3499} (\bibinfo {year} {1994})},\ \bibinfo {note} {[Erratum:
  Phys.Rev.D 53, 4111 (1996)]},\ \Eprint {https://arxiv.org/abs/hep-ph/9309298}
  {arXiv:hep-ph/9309298} \BibitemShut {NoStop}%
\bibitem [{\citenamefont {Fleischer}\ \emph {et~al.}(1993)\citenamefont
  {Fleischer}, \citenamefont {Tarasov},\ and\ \citenamefont
  {Jegerlehner}}]{Fleischer:1993ub}%
  \BibitemOpen
  \bibfield  {author} {\bibinfo {author} {\bibfnamefont {J.}~\bibnamefont
  {Fleischer}}, \bibinfo {author} {\bibfnamefont {O.~V.}\ \bibnamefont
  {Tarasov}},\ and\ \bibinfo {author} {\bibfnamefont {F.}~\bibnamefont
  {Jegerlehner}},\ }\bibfield  {title} {\bibinfo {title} {{Two loop heavy top
  corrections to the rho parameter: A Simple formula valid for arbitrary Higgs
  mass}},\ }\href {https://doi.org/10.1016/0370-2693(93)90810-5} {\bibfield
  {journal} {\bibinfo  {journal} {Phys. Lett. B}\ }\textbf {\bibinfo {volume}
  {319}},\ \bibinfo {pages} {249} (\bibinfo {year} {1993})}\BibitemShut
  {NoStop}%
\bibitem [{\citenamefont {Fleischer}\ \emph {et~al.}(1995)\citenamefont
  {Fleischer}, \citenamefont {Tarasov},\ and\ \citenamefont
  {Jegerlehner}}]{Fleischer:1994cb}%
  \BibitemOpen
  \bibfield  {author} {\bibinfo {author} {\bibfnamefont {J.}~\bibnamefont
  {Fleischer}}, \bibinfo {author} {\bibfnamefont {O.~V.}\ \bibnamefont
  {Tarasov}},\ and\ \bibinfo {author} {\bibfnamefont {F.}~\bibnamefont
  {Jegerlehner}},\ }\bibfield  {title} {\bibinfo {title} {{Two loop large top
  mass corrections to electroweak parameters: Analytic results valid for
  arbitrary Higgs mass}},\ }\href {https://doi.org/10.1103/PhysRevD.51.3820}
  {\bibfield  {journal} {\bibinfo  {journal} {Phys. Rev. D}\ }\textbf {\bibinfo
  {volume} {51}},\ \bibinfo {pages} {3820} (\bibinfo {year}
  {1995})}\BibitemShut {NoStop}%
\bibitem [{\citenamefont {Avdeev}\ \emph {et~al.}(1994)\citenamefont {Avdeev},
  \citenamefont {Fleischer}, \citenamefont {Mikhailov},\ and\ \citenamefont
  {Tarasov}}]{Avdeev:1994db}%
  \BibitemOpen
  \bibfield  {author} {\bibinfo {author} {\bibfnamefont {L.}~\bibnamefont
  {Avdeev}}, \bibinfo {author} {\bibfnamefont {J.}~\bibnamefont {Fleischer}},
  \bibinfo {author} {\bibfnamefont {S.}~\bibnamefont {Mikhailov}},\ and\
  \bibinfo {author} {\bibfnamefont {O.}~\bibnamefont {Tarasov}},\ }\bibfield
  {title} {\bibinfo {title} {{$\mathcal{0}(\alpha \alpha_s^2)$ correction to
  the electroweak $\rho$ parameter}},\ }\href
  {https://doi.org/10.1016/0370-2693(94)90573-8} {\bibfield  {journal}
  {\bibinfo  {journal} {Phys. Lett. B}\ }\textbf {\bibinfo {volume} {336}},\
  \bibinfo {pages} {560} (\bibinfo {year} {1994})},\ \bibinfo {note} {[Erratum:
  Phys.Lett.B 349, 597--598 (1995)]},\ \Eprint
  {https://arxiv.org/abs/hep-ph/9406363} {arXiv:hep-ph/9406363} \BibitemShut
  {NoStop}%
\bibitem [{\citenamefont {Chetyrkin}\ \emph
  {et~al.}(1995{\natexlab{a}})\citenamefont {Chetyrkin}, \citenamefont {Kuhn},\
  and\ \citenamefont {Steinhauser}}]{Chetyrkin:1995ix}%
  \BibitemOpen
  \bibfield  {author} {\bibinfo {author} {\bibfnamefont {K.~G.}\ \bibnamefont
  {Chetyrkin}}, \bibinfo {author} {\bibfnamefont {J.~H.}\ \bibnamefont
  {Kuhn}},\ and\ \bibinfo {author} {\bibfnamefont {M.}~\bibnamefont
  {Steinhauser}},\ }\bibfield  {title} {\bibinfo {title} {{Corrections of order
  ${\cal O}(G_F M_t^2 \alpha_s^2)$ to the $\rho$ parameter}},\ }\href
  {https://doi.org/10.1016/0370-2693(95)00380-4} {\bibfield  {journal}
  {\bibinfo  {journal} {Phys. Lett. B}\ }\textbf {\bibinfo {volume} {351}},\
  \bibinfo {pages} {331} (\bibinfo {year} {1995}{\natexlab{a}})},\ \Eprint
  {https://arxiv.org/abs/hep-ph/9502291} {arXiv:hep-ph/9502291} \BibitemShut
  {NoStop}%
\bibitem [{\citenamefont {Chetyrkin}\ \emph
  {et~al.}(1995{\natexlab{b}})\citenamefont {Chetyrkin}, \citenamefont {Kuhn},\
  and\ \citenamefont {Steinhauser}}]{Chetyrkin:1995js}%
  \BibitemOpen
  \bibfield  {author} {\bibinfo {author} {\bibfnamefont {K.~G.}\ \bibnamefont
  {Chetyrkin}}, \bibinfo {author} {\bibfnamefont {J.~H.}\ \bibnamefont
  {Kuhn}},\ and\ \bibinfo {author} {\bibfnamefont {M.}~\bibnamefont
  {Steinhauser}},\ }\bibfield  {title} {\bibinfo {title} {{QCD corrections from
  top quark to relations between electroweak parameters to order
  $alpha_s^2$}},\ }\href {https://doi.org/10.1103/PhysRevLett.75.3394}
  {\bibfield  {journal} {\bibinfo  {journal} {Phys. Rev. Lett.}\ }\textbf
  {\bibinfo {volume} {75}},\ \bibinfo {pages} {3394} (\bibinfo {year}
  {1995}{\natexlab{b}})},\ \Eprint {https://arxiv.org/abs/hep-ph/9504413}
  {arXiv:hep-ph/9504413} \BibitemShut {NoStop}%
\bibitem [{\citenamefont {Degrassi}\ \emph {et~al.}(1996)\citenamefont
  {Degrassi}, \citenamefont {Gambino},\ and\ \citenamefont
  {Vicini}}]{Degrassi:1996mg}%
  \BibitemOpen
  \bibfield  {author} {\bibinfo {author} {\bibfnamefont {G.}~\bibnamefont
  {Degrassi}}, \bibinfo {author} {\bibfnamefont {P.}~\bibnamefont {Gambino}},\
  and\ \bibinfo {author} {\bibfnamefont {A.}~\bibnamefont {Vicini}},\
  }\bibfield  {title} {\bibinfo {title} {{Two loop heavy top effects on the
  $m_Z - m_W$ interdependence}},\ }\href
  {https://doi.org/10.1016/0370-2693(96)00720-4} {\bibfield  {journal}
  {\bibinfo  {journal} {Phys. Lett. B}\ }\textbf {\bibinfo {volume} {383}},\
  \bibinfo {pages} {219} (\bibinfo {year} {1996})},\ \Eprint
  {https://arxiv.org/abs/hep-ph/9603374} {arXiv:hep-ph/9603374} \BibitemShut
  {NoStop}%
\bibitem [{\citenamefont {Degrassi}\ \emph {et~al.}(1997)\citenamefont
  {Degrassi}, \citenamefont {Gambino},\ and\ \citenamefont
  {Sirlin}}]{Degrassi:1996ps}%
  \BibitemOpen
  \bibfield  {author} {\bibinfo {author} {\bibfnamefont {G.}~\bibnamefont
  {Degrassi}}, \bibinfo {author} {\bibfnamefont {P.}~\bibnamefont {Gambino}},\
  and\ \bibinfo {author} {\bibfnamefont {A.}~\bibnamefont {Sirlin}},\
  }\bibfield  {title} {\bibinfo {title} {{Precise calculation of $M_W$, $sin^2
  \theta_W (M_Z)$, and $sin^2 \theta^{eff}_{lept}$}},\ }\href
  {https://doi.org/10.1016/S0370-2693(96)01677-2} {\bibfield  {journal}
  {\bibinfo  {journal} {Phys. Lett. B}\ }\textbf {\bibinfo {volume} {394}},\
  \bibinfo {pages} {188} (\bibinfo {year} {1997})},\ \Eprint
  {https://arxiv.org/abs/hep-ph/9611363} {arXiv:hep-ph/9611363} \BibitemShut
  {NoStop}%
\bibitem [{\citenamefont {Degrassi}\ and\ \citenamefont
  {Gambino}(2000)}]{Degrassi:1999jd}%
  \BibitemOpen
  \bibfield  {author} {\bibinfo {author} {\bibfnamefont {G.}~\bibnamefont
  {Degrassi}}\ and\ \bibinfo {author} {\bibfnamefont {P.}~\bibnamefont
  {Gambino}},\ }\bibfield  {title} {\bibinfo {title} {{Two loop heavy top
  corrections to the $Z^0$ boson partial widths}},\ }\href
  {https://doi.org/10.1016/S0550-3213(99)00729-4} {\bibfield  {journal}
  {\bibinfo  {journal} {Nucl. Phys. B}\ }\textbf {\bibinfo {volume} {567}},\
  \bibinfo {pages} {3} (\bibinfo {year} {2000})},\ \Eprint
  {https://arxiv.org/abs/hep-ph/9905472} {arXiv:hep-ph/9905472} \BibitemShut
  {NoStop}%
\bibitem [{\citenamefont {Freitas}\ \emph {et~al.}(2000)\citenamefont
  {Freitas}, \citenamefont {Hollik}, \citenamefont {Walter},\ and\
  \citenamefont {Weiglein}}]{Freitas:2000gg}%
  \BibitemOpen
  \bibfield  {author} {\bibinfo {author} {\bibfnamefont {A.}~\bibnamefont
  {Freitas}}, \bibinfo {author} {\bibfnamefont {W.}~\bibnamefont {Hollik}},
  \bibinfo {author} {\bibfnamefont {W.}~\bibnamefont {Walter}},\ and\ \bibinfo
  {author} {\bibfnamefont {G.}~\bibnamefont {Weiglein}},\ }\bibfield  {title}
  {\bibinfo {title} {{Complete fermionic two loop results for the $M_W - M_Z$
  interdependence}},\ }\href {https://doi.org/10.1016/S0370-2693(00)01263-6}
  {\bibfield  {journal} {\bibinfo  {journal} {Phys. Lett. B}\ }\textbf
  {\bibinfo {volume} {495}},\ \bibinfo {pages} {338} (\bibinfo {year}
  {2000})},\ \bibinfo {note} {[Erratum: Phys.Lett.B 570, 265 (2003)]},\ \Eprint
  {https://arxiv.org/abs/hep-ph/0007091} {arXiv:hep-ph/0007091} \BibitemShut
  {NoStop}%
\bibitem [{\citenamefont {van~der Bij}\ \emph {et~al.}(2001)\citenamefont
  {van~der Bij}, \citenamefont {Chetyrkin}, \citenamefont {Faisst},
  \citenamefont {Jikia},\ and\ \citenamefont
  {Seidensticker}}]{vanderBij:2000cg}%
  \BibitemOpen
  \bibfield  {author} {\bibinfo {author} {\bibfnamefont {J.~J.}\ \bibnamefont
  {van~der Bij}}, \bibinfo {author} {\bibfnamefont {K.~G.}\ \bibnamefont
  {Chetyrkin}}, \bibinfo {author} {\bibfnamefont {M.}~\bibnamefont {Faisst}},
  \bibinfo {author} {\bibfnamefont {G.}~\bibnamefont {Jikia}},\ and\ \bibinfo
  {author} {\bibfnamefont {T.}~\bibnamefont {Seidensticker}},\ }\bibfield
  {title} {\bibinfo {title} {{Three loop leading top mass contributions to the
  rho parameter}},\ }\href {https://doi.org/10.1016/S0370-2693(01)00002-8}
  {\bibfield  {journal} {\bibinfo  {journal} {Phys. Lett. B}\ }\textbf
  {\bibinfo {volume} {498}},\ \bibinfo {pages} {156} (\bibinfo {year}
  {2001})},\ \Eprint {https://arxiv.org/abs/hep-ph/0011373}
  {arXiv:hep-ph/0011373} \BibitemShut {NoStop}%
\bibitem [{\citenamefont {Freitas}\ \emph {et~al.}(2002)\citenamefont
  {Freitas}, \citenamefont {Hollik}, \citenamefont {Walter},\ and\
  \citenamefont {Weiglein}}]{Freitas:2002ja}%
  \BibitemOpen
  \bibfield  {author} {\bibinfo {author} {\bibfnamefont {A.}~\bibnamefont
  {Freitas}}, \bibinfo {author} {\bibfnamefont {W.}~\bibnamefont {Hollik}},
  \bibinfo {author} {\bibfnamefont {W.}~\bibnamefont {Walter}},\ and\ \bibinfo
  {author} {\bibfnamefont {G.}~\bibnamefont {Weiglein}},\ }\bibfield  {title}
  {\bibinfo {title} {{Electroweak two loop corrections to the $M_W-M_Z$ mass
  correlation in the standard model}},\ }\href
  {https://doi.org/10.1016/S0550-3213(02)00243-2} {\bibfield  {journal}
  {\bibinfo  {journal} {Nucl. Phys. B}\ }\textbf {\bibinfo {volume} {632}},\
  \bibinfo {pages} {189} (\bibinfo {year} {2002})},\ \bibinfo {note} {[Erratum:
  Nucl.Phys.B 666, 305--307 (2003)]},\ \Eprint
  {https://arxiv.org/abs/hep-ph/0202131} {arXiv:hep-ph/0202131} \BibitemShut
  {NoStop}%
\bibitem [{\citenamefont {Awramik}\ and\ \citenamefont
  {Czakon}(2002)}]{Awramik:2002wn}%
  \BibitemOpen
  \bibfield  {author} {\bibinfo {author} {\bibfnamefont {M.}~\bibnamefont
  {Awramik}}\ and\ \bibinfo {author} {\bibfnamefont {M.}~\bibnamefont
  {Czakon}},\ }\bibfield  {title} {\bibinfo {title} {{Complete two loop bosonic
  contributions to the muon lifetime in the standard model}},\ }\href
  {https://doi.org/10.1103/PhysRevLett.89.241801} {\bibfield  {journal}
  {\bibinfo  {journal} {Phys. Rev. Lett.}\ }\textbf {\bibinfo {volume} {89}},\
  \bibinfo {pages} {241801} (\bibinfo {year} {2002})},\ \Eprint
  {https://arxiv.org/abs/hep-ph/0208113} {arXiv:hep-ph/0208113} \BibitemShut
  {NoStop}%
\bibitem [{\citenamefont {Onishchenko}\ and\ \citenamefont
  {Veretin}(2003)}]{Onishchenko:2002ve}%
  \BibitemOpen
  \bibfield  {author} {\bibinfo {author} {\bibfnamefont {A.}~\bibnamefont
  {Onishchenko}}\ and\ \bibinfo {author} {\bibfnamefont {O.}~\bibnamefont
  {Veretin}},\ }\bibfield  {title} {\bibinfo {title} {{Two loop bosonic
  electroweak corrections to the muon lifetime and $M_Z - M_W$
  interdependence}},\ }\href {https://doi.org/10.1016/S0370-2693(02)03004-6}
  {\bibfield  {journal} {\bibinfo  {journal} {Phys. Lett. B}\ }\textbf
  {\bibinfo {volume} {551}},\ \bibinfo {pages} {111} (\bibinfo {year}
  {2003})},\ \Eprint {https://arxiv.org/abs/hep-ph/0209010}
  {arXiv:hep-ph/0209010} \BibitemShut {NoStop}%
\bibitem [{\citenamefont {Awramik}\ \emph {et~al.}(2003)\citenamefont
  {Awramik}, \citenamefont {Czakon}, \citenamefont {Onishchenko},\ and\
  \citenamefont {Veretin}}]{Awramik:2002vu}%
  \BibitemOpen
  \bibfield  {author} {\bibinfo {author} {\bibfnamefont {M.}~\bibnamefont
  {Awramik}}, \bibinfo {author} {\bibfnamefont {M.}~\bibnamefont {Czakon}},
  \bibinfo {author} {\bibfnamefont {A.}~\bibnamefont {Onishchenko}},\ and\
  \bibinfo {author} {\bibfnamefont {O.}~\bibnamefont {Veretin}},\ }\bibfield
  {title} {\bibinfo {title} {{Bosonic corrections to $\Delta r$ at the two loop
  level}},\ }\href {https://doi.org/10.1103/PhysRevD.68.053004} {\bibfield
  {journal} {\bibinfo  {journal} {Phys. Rev. D}\ }\textbf {\bibinfo {volume}
  {68}},\ \bibinfo {pages} {053004} (\bibinfo {year} {2003})},\ \Eprint
  {https://arxiv.org/abs/hep-ph/0209084} {arXiv:hep-ph/0209084} \BibitemShut
  {NoStop}%
\bibitem [{\citenamefont {Awramik}\ and\ \citenamefont
  {Czakon}(2003{\natexlab{a}})}]{Awramik:2002wv}%
  \BibitemOpen
  \bibfield  {author} {\bibinfo {author} {\bibfnamefont {M.}~\bibnamefont
  {Awramik}}\ and\ \bibinfo {author} {\bibfnamefont {M.}~\bibnamefont
  {Czakon}},\ }\bibfield  {title} {\bibinfo {title} {{Two loop electroweak
  bosonic corrections to the muon decay lifetime}},\ }\href
  {https://doi.org/10.1016/S0920-5632(03)80177-9} {\bibfield  {journal}
  {\bibinfo  {journal} {Nucl. Phys. B Proc. Suppl.}\ }\textbf {\bibinfo
  {volume} {116}},\ \bibinfo {pages} {238} (\bibinfo {year}
  {2003}{\natexlab{a}})},\ \Eprint {https://arxiv.org/abs/hep-ph/0211041}
  {arXiv:hep-ph/0211041} \BibitemShut {NoStop}%
\bibitem [{\citenamefont {Awramik}\ and\ \citenamefont
  {Czakon}(2003{\natexlab{b}})}]{Awramik:2003ee}%
  \BibitemOpen
  \bibfield  {author} {\bibinfo {author} {\bibfnamefont {M.}~\bibnamefont
  {Awramik}}\ and\ \bibinfo {author} {\bibfnamefont {M.}~\bibnamefont
  {Czakon}},\ }\bibfield  {title} {\bibinfo {title} {{Complete two loop
  electroweak contributions to the muon lifetime in the standard model}},\
  }\href {https://doi.org/10.1016/j.physletb.2003.06.007} {\bibfield  {journal}
  {\bibinfo  {journal} {Phys. Lett. B}\ }\textbf {\bibinfo {volume} {568}},\
  \bibinfo {pages} {48} (\bibinfo {year} {2003}{\natexlab{b}})},\ \Eprint
  {https://arxiv.org/abs/hep-ph/0305248} {arXiv:hep-ph/0305248} \BibitemShut
  {NoStop}%
\bibitem [{\citenamefont {Awramik}\ \emph {et~al.}(2004)\citenamefont
  {Awramik}, \citenamefont {Czakon}, \citenamefont {Freitas},\ and\
  \citenamefont {Weiglein}}]{Awramik:2003rn}%
  \BibitemOpen
  \bibfield  {author} {\bibinfo {author} {\bibfnamefont {M.}~\bibnamefont
  {Awramik}}, \bibinfo {author} {\bibfnamefont {M.}~\bibnamefont {Czakon}},
  \bibinfo {author} {\bibfnamefont {A.}~\bibnamefont {Freitas}},\ and\ \bibinfo
  {author} {\bibfnamefont {G.}~\bibnamefont {Weiglein}},\ }\bibfield  {title}
  {\bibinfo {title} {{Precise prediction for the $W$ boson mass in the standard
  model}},\ }\href {https://doi.org/10.1103/PhysRevD.69.053006} {\bibfield
  {journal} {\bibinfo  {journal} {Phys. Rev. D}\ }\textbf {\bibinfo {volume}
  {69}},\ \bibinfo {pages} {053006} (\bibinfo {year} {2004})},\ \Eprint
  {https://arxiv.org/abs/hep-ph/0311148} {arXiv:hep-ph/0311148} \BibitemShut
  {NoStop}%
\bibitem [{\citenamefont {Faisst}\ \emph {et~al.}(2003)\citenamefont {Faisst},
  \citenamefont {Kuhn}, \citenamefont {Seidensticker},\ and\ \citenamefont
  {Veretin}}]{Faisst:2003px}%
  \BibitemOpen
  \bibfield  {author} {\bibinfo {author} {\bibfnamefont {M.}~\bibnamefont
  {Faisst}}, \bibinfo {author} {\bibfnamefont {J.~H.}\ \bibnamefont {Kuhn}},
  \bibinfo {author} {\bibfnamefont {T.}~\bibnamefont {Seidensticker}},\ and\
  \bibinfo {author} {\bibfnamefont {O.}~\bibnamefont {Veretin}},\ }\bibfield
  {title} {\bibinfo {title} {{Three loop top quark contributions to the $\rho$
  parameter}},\ }\href {https://doi.org/10.1016/S0550-3213(03)00450-4}
  {\bibfield  {journal} {\bibinfo  {journal} {Nucl. Phys. B}\ }\textbf
  {\bibinfo {volume} {665}},\ \bibinfo {pages} {649} (\bibinfo {year}
  {2003})},\ \Eprint {https://arxiv.org/abs/hep-ph/0302275}
  {arXiv:hep-ph/0302275} \BibitemShut {NoStop}%
\bibitem [{\citenamefont {Dubovyk}\ \emph {et~al.}(2016)\citenamefont
  {Dubovyk}, \citenamefont {Freitas}, \citenamefont {Gluza}, \citenamefont
  {Riemann},\ and\ \citenamefont {Usovitsch}}]{Dubovyk:2016aqv}%
  \BibitemOpen
  \bibfield  {author} {\bibinfo {author} {\bibfnamefont {I.}~\bibnamefont
  {Dubovyk}}, \bibinfo {author} {\bibfnamefont {A.}~\bibnamefont {Freitas}},
  \bibinfo {author} {\bibfnamefont {J.}~\bibnamefont {Gluza}}, \bibinfo
  {author} {\bibfnamefont {T.}~\bibnamefont {Riemann}},\ and\ \bibinfo {author}
  {\bibfnamefont {J.}~\bibnamefont {Usovitsch}},\ }\bibfield  {title} {\bibinfo
  {title} {{The two-loop electroweak bosonic corrections to
  $\sin^2\theta^\textrm{b}_\textrm{eff}$}},\ }\href
  {https://doi.org/10.1016/j.physletb.2016.09.012} {\bibfield  {journal}
  {\bibinfo  {journal} {Phys. Lett. B}\ }\textbf {\bibinfo {volume} {762}},\
  \bibinfo {pages} {184} (\bibinfo {year} {2016})},\ \Eprint
  {https://arxiv.org/abs/1607.08375} {arXiv:1607.08375 [hep-ph]} \BibitemShut
  {NoStop}%
\bibitem [{\citenamefont {Dubovyk}\ \emph {et~al.}(2018)\citenamefont
  {Dubovyk}, \citenamefont {Freitas}, \citenamefont {Gluza}, \citenamefont
  {Riemann},\ and\ \citenamefont {Usovitsch}}]{Dubovyk:2018rlg}%
  \BibitemOpen
  \bibfield  {author} {\bibinfo {author} {\bibfnamefont {I.}~\bibnamefont
  {Dubovyk}}, \bibinfo {author} {\bibfnamefont {A.}~\bibnamefont {Freitas}},
  \bibinfo {author} {\bibfnamefont {J.}~\bibnamefont {Gluza}}, \bibinfo
  {author} {\bibfnamefont {T.}~\bibnamefont {Riemann}},\ and\ \bibinfo {author}
  {\bibfnamefont {J.}~\bibnamefont {Usovitsch}},\ }\bibfield  {title} {\bibinfo
  {title} {{Complete electroweak two-loop corrections to $Z$ boson production
  and decay}},\ }\href {https://doi.org/10.1016/j.physletb.2018.06.037}
  {\bibfield  {journal} {\bibinfo  {journal} {Phys. Lett. B}\ }\textbf
  {\bibinfo {volume} {783}},\ \bibinfo {pages} {86} (\bibinfo {year} {2018})},\
  \Eprint {https://arxiv.org/abs/1804.10236} {arXiv:1804.10236 [hep-ph]}
  \BibitemShut {NoStop}%
\bibitem [{\citenamefont {Barbieri}(2007)}]{Barbieri:2007gi}%
  \BibitemOpen
  \bibfield  {author} {\bibinfo {author} {\bibfnamefont {R.}~\bibnamefont
  {Barbieri}},\ }\href@noop {} {\emph {\bibinfo {title} {{Ten Lectures on the
  ElectroWeak Interactions}}}}\ (\bibinfo  {publisher} {Scuola Normale
  Superiore},\ \bibinfo {year} {2007})\ \Eprint
  {https://arxiv.org/abs/0706.0684} {arXiv:0706.0684 [hep-ph]} \BibitemShut
  {NoStop}%
\bibitem [{\citenamefont {Ellis}\ \emph {et~al.}(2021)\citenamefont {Ellis},
  \citenamefont {Madigan}, \citenamefont {Mimasu}, \citenamefont {Sanz},\ and\
  \citenamefont {You}}]{Ellis:2020unq}%
  \BibitemOpen
  \bibfield  {author} {\bibinfo {author} {\bibfnamefont {J.}~\bibnamefont
  {Ellis}}, \bibinfo {author} {\bibfnamefont {M.}~\bibnamefont {Madigan}},
  \bibinfo {author} {\bibfnamefont {K.}~\bibnamefont {Mimasu}}, \bibinfo
  {author} {\bibfnamefont {V.}~\bibnamefont {Sanz}},\ and\ \bibinfo {author}
  {\bibfnamefont {T.}~\bibnamefont {You}},\ }\bibfield  {title} {\bibinfo
  {title} {{Top, Higgs, Diboson and Electroweak Fit to the Standard Model
  Effective Field Theory}},\ }\href {https://doi.org/10.1007/JHEP04(2021)279}
  {\bibfield  {journal} {\bibinfo  {journal} {JHEP}\ }\textbf {\bibinfo
  {volume} {04}},\ \bibinfo {pages} {279}},\ \Eprint
  {https://arxiv.org/abs/2012.02779} {arXiv:2012.02779 [hep-ph]} \BibitemShut
  {NoStop}%
\bibitem [{\citenamefont {De~Blas}\ \emph {et~al.}(2019)\citenamefont
  {De~Blas}, \citenamefont {Durieux}, \citenamefont {Grojean}, \citenamefont
  {Gu},\ and\ \citenamefont {Paul}}]{DeBlas:2019qco}%
  \BibitemOpen
  \bibfield  {author} {\bibinfo {author} {\bibfnamefont {J.}~\bibnamefont
  {De~Blas}}, \bibinfo {author} {\bibfnamefont {G.}~\bibnamefont {Durieux}},
  \bibinfo {author} {\bibfnamefont {C.}~\bibnamefont {Grojean}}, \bibinfo
  {author} {\bibfnamefont {J.}~\bibnamefont {Gu}},\ and\ \bibinfo {author}
  {\bibfnamefont {A.}~\bibnamefont {Paul}},\ }\bibfield  {title} {\bibinfo
  {title} {{On the future of Higgs, electroweak and diboson measurements at
  lepton colliders}},\ }\href {https://doi.org/10.1007/JHEP12(2019)117}
  {\bibfield  {journal} {\bibinfo  {journal} {JHEP}\ }\textbf {\bibinfo
  {volume} {12}},\ \bibinfo {pages} {117}},\ \Eprint
  {https://arxiv.org/abs/1907.04311} {arXiv:1907.04311 [hep-ph]} \BibitemShut
  {NoStop}%
\bibitem [{\citenamefont {de~Blas}\ \emph {et~al.}(2020)\citenamefont {de~Blas}
  \emph {et~al.}}]{deBlas:2019rxi}%
  \BibitemOpen
  \bibfield  {author} {\bibinfo {author} {\bibfnamefont {J.}~\bibnamefont
  {de~Blas}} \emph {et~al.},\ }\bibfield  {title} {\bibinfo {title} {{Higgs
  Boson Studies at Future Particle Colliders}},\ }\href
  {https://doi.org/10.1007/JHEP01(2020)139} {\bibfield  {journal} {\bibinfo
  {journal} {JHEP}\ }\textbf {\bibinfo {volume} {01}},\ \bibinfo {pages}
  {139}},\ \Eprint {https://arxiv.org/abs/1905.03764} {arXiv:1905.03764
  [hep-ph]} \BibitemShut {NoStop}%
\bibitem [{\citenamefont {Ellis}\ \emph {et~al.}(2018)\citenamefont {Ellis},
  \citenamefont {Murphy}, \citenamefont {Sanz},\ and\ \citenamefont
  {You}}]{Ellis:2018gqa}%
  \BibitemOpen
  \bibfield  {author} {\bibinfo {author} {\bibfnamefont {J.}~\bibnamefont
  {Ellis}}, \bibinfo {author} {\bibfnamefont {C.~W.}\ \bibnamefont {Murphy}},
  \bibinfo {author} {\bibfnamefont {V.}~\bibnamefont {Sanz}},\ and\ \bibinfo
  {author} {\bibfnamefont {T.}~\bibnamefont {You}},\ }\bibfield  {title}
  {\bibinfo {title} {{Updated Global SMEFT Fit to Higgs, Diboson and
  Electroweak Data}},\ }\href {https://doi.org/10.1007/JHEP06(2018)146}
  {\bibfield  {journal} {\bibinfo  {journal} {JHEP}\ }\textbf {\bibinfo
  {volume} {06}},\ \bibinfo {pages} {146}},\ \Eprint
  {https://arxiv.org/abs/1803.03252} {arXiv:1803.03252 [hep-ph]} \BibitemShut
  {NoStop}%
\bibitem [{Note1()}]{Note1}%
  \BibitemOpen
  \bibinfo {note} {As recently discussed in Ref.~\cite {Kribs:2020jgn}, a
  general characterization of UV breaking effects of $SU(2)_{L} \times
  SU(2)_{R}$ may require some care. In this study, we simply regard these
  effects as those originally highlighted in Ref.~\cite {Peskin:1990zt} in the
  limit of zero momentum.}\BibitemShut {Stop}%
\bibitem [{\citenamefont {Alasfar}\ \emph {et~al.}(2020)\citenamefont
  {Alasfar}, \citenamefont {Azatov}, \citenamefont {de~Blas}, \citenamefont
  {Paul},\ and\ \citenamefont {Valli}}]{Alasfar:2020mne}%
  \BibitemOpen
  \bibfield  {author} {\bibinfo {author} {\bibfnamefont {L.}~\bibnamefont
  {Alasfar}}, \bibinfo {author} {\bibfnamefont {A.}~\bibnamefont {Azatov}},
  \bibinfo {author} {\bibfnamefont {J.}~\bibnamefont {de~Blas}}, \bibinfo
  {author} {\bibfnamefont {A.}~\bibnamefont {Paul}},\ and\ \bibinfo {author}
  {\bibfnamefont {M.}~\bibnamefont {Valli}},\ }\bibfield  {title} {\bibinfo
  {title} {{$B$ anomalies under the lens of electroweak precision}},\ }\href
  {https://doi.org/10.1007/JHEP12(2020)016} {\bibfield  {journal} {\bibinfo
  {journal} {JHEP}\ }\textbf {\bibinfo {volume} {12}},\ \bibinfo {pages}
  {016}},\ \Eprint {https://arxiv.org/abs/2007.04400} {arXiv:2007.04400
  [hep-ph]} \BibitemShut {NoStop}%
\bibitem [{\citenamefont {Grzadkowski}\ \emph {et~al.}(2010)\citenamefont
  {Grzadkowski}, \citenamefont {Iskrzynski}, \citenamefont {Misiak},\ and\
  \citenamefont {Rosiek}}]{Grzadkowski:2010es}%
  \BibitemOpen
  \bibfield  {author} {\bibinfo {author} {\bibfnamefont {B.}~\bibnamefont
  {Grzadkowski}}, \bibinfo {author} {\bibfnamefont {M.}~\bibnamefont
  {Iskrzynski}}, \bibinfo {author} {\bibfnamefont {M.}~\bibnamefont {Misiak}},\
  and\ \bibinfo {author} {\bibfnamefont {J.}~\bibnamefont {Rosiek}},\
  }\bibfield  {title} {\bibinfo {title} {{Dimension-Six Terms in the Standard
  Model Lagrangian}},\ }\href {https://doi.org/10.1007/JHEP10(2010)085}
  {\bibfield  {journal} {\bibinfo  {journal} {JHEP}\ }\textbf {\bibinfo
  {volume} {10}},\ \bibinfo {pages} {085}},\ \Eprint
  {https://arxiv.org/abs/1008.4884} {arXiv:1008.4884 [hep-ph]} \BibitemShut
  {NoStop}%
\bibitem [{\citenamefont {Berthier}\ and\ \citenamefont
  {Trott}(2015)}]{Berthier:2015oma}%
  \BibitemOpen
  \bibfield  {author} {\bibinfo {author} {\bibfnamefont {L.}~\bibnamefont
  {Berthier}}\ and\ \bibinfo {author} {\bibfnamefont {M.}~\bibnamefont
  {Trott}},\ }\bibfield  {title} {\bibinfo {title} {{Towards consistent
  Electroweak Precision Data constraints in the SMEFT}},\ }\href
  {https://doi.org/10.1007/JHEP05(2015)024} {\bibfield  {journal} {\bibinfo
  {journal} {JHEP}\ }\textbf {\bibinfo {volume} {05}},\ \bibinfo {pages}
  {024}},\ \Eprint {https://arxiv.org/abs/1502.02570} {arXiv:1502.02570
  [hep-ph]} \BibitemShut {NoStop}%
\bibitem [{\citenamefont {Bj\o{}rn}\ and\ \citenamefont
  {Trott}(2016)}]{Bjorn:2016zlr}%
  \BibitemOpen
  \bibfield  {author} {\bibinfo {author} {\bibfnamefont {M.}~\bibnamefont
  {Bj\o{}rn}}\ and\ \bibinfo {author} {\bibfnamefont {M.}~\bibnamefont
  {Trott}},\ }\bibfield  {title} {\bibinfo {title} {{Interpreting $W$ mass
  measurements in the SMEFT}},\ }\href
  {https://doi.org/10.1016/j.physletb.2016.10.003} {\bibfield  {journal}
  {\bibinfo  {journal} {Phys. Lett. B}\ }\textbf {\bibinfo {volume} {762}},\
  \bibinfo {pages} {426} (\bibinfo {year} {2016})},\ \Eprint
  {https://arxiv.org/abs/1606.06502} {arXiv:1606.06502 [hep-ph]} \BibitemShut
  {NoStop}%
\bibitem [{\citenamefont {Efrati}\ \emph {et~al.}(2015)\citenamefont {Efrati},
  \citenamefont {Falkowski},\ and\ \citenamefont {Soreq}}]{Efrati:2015eaa}%
  \BibitemOpen
  \bibfield  {author} {\bibinfo {author} {\bibfnamefont {A.}~\bibnamefont
  {Efrati}}, \bibinfo {author} {\bibfnamefont {A.}~\bibnamefont {Falkowski}},\
  and\ \bibinfo {author} {\bibfnamefont {Y.}~\bibnamefont {Soreq}},\ }\bibfield
   {title} {\bibinfo {title} {{Electroweak constraints on flavorful effective
  theories}},\ }\href {https://doi.org/10.1007/JHEP07(2015)018} {\bibfield
  {journal} {\bibinfo  {journal} {JHEP}\ }\textbf {\bibinfo {volume} {07}},\
  \bibinfo {pages} {018}},\ \Eprint {https://arxiv.org/abs/1503.07872}
  {arXiv:1503.07872 [hep-ph]} \BibitemShut {NoStop}%
\bibitem [{\citenamefont {Peskin}\ and\ \citenamefont
  {Takeuchi}(1990)}]{Peskin:1990zt}%
  \BibitemOpen
  \bibfield  {author} {\bibinfo {author} {\bibfnamefont {M.~E.}\ \bibnamefont
  {Peskin}}\ and\ \bibinfo {author} {\bibfnamefont {T.}~\bibnamefont
  {Takeuchi}},\ }\bibfield  {title} {\bibinfo {title} {{A New constraint on a
  strongly interacting Higgs sector}},\ }\href
  {https://doi.org/10.1103/PhysRevLett.65.964} {\bibfield  {journal} {\bibinfo
  {journal} {Phys. Rev. Lett.}\ }\textbf {\bibinfo {volume} {65}},\ \bibinfo
  {pages} {964} (\bibinfo {year} {1990})}\BibitemShut {NoStop}%
\bibitem [{\citenamefont {Peskin}\ and\ \citenamefont
  {Takeuchi}(1992)}]{Peskin:1991sw}%
  \BibitemOpen
  \bibfield  {author} {\bibinfo {author} {\bibfnamefont {M.~E.}\ \bibnamefont
  {Peskin}}\ and\ \bibinfo {author} {\bibfnamefont {T.}~\bibnamefont
  {Takeuchi}},\ }\bibfield  {title} {\bibinfo {title} {{Estimation of oblique
  electroweak corrections}},\ }\href {https://doi.org/10.1103/PhysRevD.46.381}
  {\bibfield  {journal} {\bibinfo  {journal} {Phys. Rev. D}\ }\textbf {\bibinfo
  {volume} {46}},\ \bibinfo {pages} {381} (\bibinfo {year} {1992})}\BibitemShut
  {NoStop}%
\bibitem [{\citenamefont {Altarelli}\ and\ \citenamefont
  {Barbieri}(1991)}]{Altarelli:1990zd}%
  \BibitemOpen
  \bibfield  {author} {\bibinfo {author} {\bibfnamefont {G.}~\bibnamefont
  {Altarelli}}\ and\ \bibinfo {author} {\bibfnamefont {R.}~\bibnamefont
  {Barbieri}},\ }\bibfield  {title} {\bibinfo {title} {{Vacuum polarization
  effects of new physics on electroweak processes}},\ }\href
  {https://doi.org/10.1016/0370-2693(91)91378-9} {\bibfield  {journal}
  {\bibinfo  {journal} {Phys. Lett. B}\ }\textbf {\bibinfo {volume} {253}},\
  \bibinfo {pages} {161} (\bibinfo {year} {1991})}\BibitemShut {NoStop}%
\bibitem [{\citenamefont {Altarelli}\ \emph {et~al.}(1992)\citenamefont
  {Altarelli}, \citenamefont {Barbieri},\ and\ \citenamefont
  {Jadach}}]{Altarelli:1991fk}%
  \BibitemOpen
  \bibfield  {author} {\bibinfo {author} {\bibfnamefont {G.}~\bibnamefont
  {Altarelli}}, \bibinfo {author} {\bibfnamefont {R.}~\bibnamefont
  {Barbieri}},\ and\ \bibinfo {author} {\bibfnamefont {S.}~\bibnamefont
  {Jadach}},\ }\bibfield  {title} {\bibinfo {title} {{Toward a model
  independent analysis of electroweak data}},\ }\href
  {https://doi.org/10.1016/0550-3213(92)90376-M} {\bibfield  {journal}
  {\bibinfo  {journal} {Nucl. Phys. B}\ }\textbf {\bibinfo {volume} {369}},\
  \bibinfo {pages} {3} (\bibinfo {year} {1992})},\ \bibinfo {note} {[Erratum:
  Nucl.Phys.B 376, 444 (1992)]}\BibitemShut {NoStop}%
\bibitem [{\citenamefont {Altarelli}\ \emph {et~al.}(1993)\citenamefont
  {Altarelli}, \citenamefont {Barbieri},\ and\ \citenamefont
  {Caravaglios}}]{Altarelli:1993sz}%
  \BibitemOpen
  \bibfield  {author} {\bibinfo {author} {\bibfnamefont {G.}~\bibnamefont
  {Altarelli}}, \bibinfo {author} {\bibfnamefont {R.}~\bibnamefont
  {Barbieri}},\ and\ \bibinfo {author} {\bibfnamefont {F.}~\bibnamefont
  {Caravaglios}},\ }\bibfield  {title} {\bibinfo {title} {{Nonstandard analysis
  of electroweak precision data}},\ }\href
  {https://doi.org/10.1016/0550-3213(93)90424-N} {\bibfield  {journal}
  {\bibinfo  {journal} {Nucl. Phys. B}\ }\textbf {\bibinfo {volume} {405}},\
  \bibinfo {pages} {3} (\bibinfo {year} {1993})}\BibitemShut {NoStop}%
\bibitem [{\citenamefont {{Grinstein}}\ and\ \citenamefont
  {{Wise}}(1991)}]{1991PhLB..265..326G}%
  \BibitemOpen
  \bibfield  {author} {\bibinfo {author} {\bibfnamefont {B.}~\bibnamefont
  {{Grinstein}}}\ and\ \bibinfo {author} {\bibfnamefont {M.~B.}\ \bibnamefont
  {{Wise}}},\ }\bibfield  {title} {\bibinfo {title} {{Operator analysis for
  precision electroweak physics}},\ }\href
  {https://doi.org/10.1016/0370-2693(91)90061-T} {\bibfield  {journal}
  {\bibinfo  {journal} {Physics Letters B}\ }\textbf {\bibinfo {volume}
  {265}},\ \bibinfo {pages} {326} (\bibinfo {year} {1991})}\BibitemShut
  {NoStop}%
\bibitem [{\citenamefont {Aaltonen}\ \emph {et~al.}(2007)\citenamefont
  {Aaltonen} \emph {et~al.}}]{CDF:2007cmy}%
  \BibitemOpen
  \bibfield  {author} {\bibinfo {author} {\bibfnamefont {T.}~\bibnamefont
  {Aaltonen}} \emph {et~al.} (\bibinfo {collaboration} {CDF}),\ }\bibfield
  {title} {\bibinfo {title} {{First measurement of the $W$ boson mass in Run II
  of the Tevatron}},\ }\href {https://doi.org/10.1103/PhysRevLett.99.151801}
  {\bibfield  {journal} {\bibinfo  {journal} {Phys. Rev. Lett.}\ }\textbf
  {\bibinfo {volume} {99}},\ \bibinfo {pages} {151801} (\bibinfo {year}
  {2007})},\ \Eprint {https://arxiv.org/abs/0707.0085} {arXiv:0707.0085
  [hep-ex]} \BibitemShut {NoStop}%
\bibitem [{\citenamefont {De~Blas}\ \emph {et~al.}(2020)\citenamefont {De~Blas}
  \emph {et~al.}}]{DeBlas:2019ehy}%
  \BibitemOpen
  \bibfield  {author} {\bibinfo {author} {\bibfnamefont {J.}~\bibnamefont
  {De~Blas}} \emph {et~al.},\ }\bibfield  {title} {\bibinfo {title}
  {{$\texttt{HEPfit}$: a code for the combination of indirect and direct
  constraints on high energy physics models}},\ }\href
  {https://doi.org/10.1140/epjc/s10052-020-7904-z} {\bibfield  {journal}
  {\bibinfo  {journal} {Eur. Phys. J. C}\ }\textbf {\bibinfo {volume} {80}},\
  \bibinfo {pages} {456} (\bibinfo {year} {2020})},\ \Eprint
  {https://arxiv.org/abs/1910.14012} {arXiv:1910.14012 [hep-ph]} \BibitemShut
  {NoStop}%
\bibitem [{\citenamefont {de~Blas}\ \emph {et~al.}(2021)\citenamefont
  {de~Blas}, \citenamefont {Ciuchini}, \citenamefont {Franco}, \citenamefont
  {Goncalves}, \citenamefont {Mishima}, \citenamefont {Pierini}, \citenamefont
  {Reina},\ and\ \citenamefont {Silvestrini}}]{deBlas:2021wap}%
  \BibitemOpen
  \bibfield  {author} {\bibinfo {author} {\bibfnamefont {J.}~\bibnamefont
  {de~Blas}}, \bibinfo {author} {\bibfnamefont {M.}~\bibnamefont {Ciuchini}},
  \bibinfo {author} {\bibfnamefont {E.}~\bibnamefont {Franco}}, \bibinfo
  {author} {\bibfnamefont {A.}~\bibnamefont {Goncalves}}, \bibinfo {author}
  {\bibfnamefont {S.}~\bibnamefont {Mishima}}, \bibinfo {author} {\bibfnamefont
  {M.}~\bibnamefont {Pierini}}, \bibinfo {author} {\bibfnamefont
  {L.}~\bibnamefont {Reina}},\ and\ \bibinfo {author} {\bibfnamefont
  {L.}~\bibnamefont {Silvestrini}},\ }\bibfield  {title} {\bibinfo {title}
  {{Global analysis of electroweak data in the Standard Model}},\ }\href@noop
  {} {\  (\bibinfo {year} {2021})},\ \Eprint {https://arxiv.org/abs/2112.07274}
  {arXiv:2112.07274 [hep-ph]} \BibitemShut {NoStop}%
\bibitem [{\citenamefont {Ando}(2011)}]{IC}%
  \BibitemOpen
  \bibfield  {author} {\bibinfo {author} {\bibfnamefont {T.}~\bibnamefont
  {Ando}},\ }\bibfield  {title} {\bibinfo {title} {Predictive bayesian model
  selection},\ }\href {https://doi.org/10.1080/01966324.2011.10737798}
  {\bibfield  {journal} {\bibinfo  {journal} {American Journal of Mathematical
  and Management Sciences}\ }\textbf {\bibinfo {volume} {31}},\ \bibinfo
  {pages} {13} (\bibinfo {year} {2011})},\ \bibinfo {note}
  {\url{http://dx.doi.org/10.1080/01966324.2011.10737798}}\BibitemShut
  {NoStop}%
\bibitem [{\citenamefont {Kass}\ and\ \citenamefont
  {Raftery}(1995)}]{BayesFactors}%
  \BibitemOpen
  \bibfield  {author} {\bibinfo {author} {\bibfnamefont {R.~E.}\ \bibnamefont
  {Kass}}\ and\ \bibinfo {author} {\bibfnamefont {A.~E.}\ \bibnamefont
  {Raftery}},\ }\bibfield  {title} {\bibinfo {title} {Bayes factors},\ }\href
  {https://doi.org/10.1080/01621459.1995.10476572} {\bibfield  {journal}
  {\bibinfo  {journal} {Journal of the American Statistical Association}\
  }\textbf {\bibinfo {volume} {90}},\ \bibinfo {pages} {773} (\bibinfo {year}
  {1995})},\ \bibinfo {note}
  {\url{http://dx.doi.org/10.1080/01621459.1995.10476572}}\BibitemShut
  {NoStop}%
\bibitem [{\citenamefont {Passera}\ \emph {et~al.}(2008)\citenamefont
  {Passera}, \citenamefont {Marciano},\ and\ \citenamefont
  {Sirlin}}]{Passera:2008jk}%
  \BibitemOpen
  \bibfield  {author} {\bibinfo {author} {\bibfnamefont {M.}~\bibnamefont
  {Passera}}, \bibinfo {author} {\bibfnamefont {W.~J.}\ \bibnamefont
  {Marciano}},\ and\ \bibinfo {author} {\bibfnamefont {A.}~\bibnamefont
  {Sirlin}},\ }\bibfield  {title} {\bibinfo {title} {{The Muon $g-2$ and the
  bounds on the Higgs boson mass}},\ }\href
  {https://doi.org/10.1103/PhysRevD.78.013009} {\bibfield  {journal} {\bibinfo
  {journal} {Phys. Rev. D}\ }\textbf {\bibinfo {volume} {78}},\ \bibinfo
  {pages} {013009} (\bibinfo {year} {2008})},\ \Eprint
  {https://arxiv.org/abs/0804.1142} {arXiv:0804.1142 [hep-ph]} \BibitemShut
  {NoStop}%
\bibitem [{\citenamefont {Abi}\ \emph {et~al.}(2021)\citenamefont {Abi} \emph
  {et~al.}}]{Muong-2:2021ojo}%
  \BibitemOpen
  \bibfield  {author} {\bibinfo {author} {\bibfnamefont {B.}~\bibnamefont
  {Abi}} \emph {et~al.} (\bibinfo {collaboration} {Muon g-2}),\ }\bibfield
  {title} {\bibinfo {title} {{Measurement of the Positive Muon Anomalous
  Magnetic Moment to 0.46 ppm}},\ }\href
  {https://doi.org/10.1103/PhysRevLett.126.141801} {\bibfield  {journal}
  {\bibinfo  {journal} {Phys. Rev. Lett.}\ }\textbf {\bibinfo {volume} {126}},\
  \bibinfo {pages} {141801} (\bibinfo {year} {2021})},\ \Eprint
  {https://arxiv.org/abs/2104.03281} {arXiv:2104.03281 [hep-ex]} \BibitemShut
  {NoStop}%
\bibitem [{\citenamefont {Bennett}\ \emph {et~al.}(2006)\citenamefont {Bennett}
  \emph {et~al.}}]{Muong-2:2006rrc}%
  \BibitemOpen
  \bibfield  {author} {\bibinfo {author} {\bibfnamefont {G.~W.}\ \bibnamefont
  {Bennett}} \emph {et~al.} (\bibinfo {collaboration} {Muon g-2}),\ }\bibfield
  {title} {\bibinfo {title} {{Final Report of the Muon E821 Anomalous Magnetic
  Moment Measurement at BNL}},\ }\href
  {https://doi.org/10.1103/PhysRevD.73.072003} {\bibfield  {journal} {\bibinfo
  {journal} {Phys. Rev. D}\ }\textbf {\bibinfo {volume} {73}},\ \bibinfo
  {pages} {072003} (\bibinfo {year} {2006})},\ \Eprint
  {https://arxiv.org/abs/hep-ex/0602035} {arXiv:hep-ex/0602035} \BibitemShut
  {NoStop}%
\bibitem [{\citenamefont {Aoyama}\ \emph {et~al.}(2020)\citenamefont {Aoyama}
  \emph {et~al.}}]{Aoyama:2020ynm}%
  \BibitemOpen
  \bibfield  {author} {\bibinfo {author} {\bibfnamefont {T.}~\bibnamefont
  {Aoyama}} \emph {et~al.},\ }\bibfield  {title} {\bibinfo {title} {{The
  anomalous magnetic moment of the muon in the Standard Model}},\ }\href
  {https://doi.org/10.1016/j.physrep.2020.07.006} {\bibfield  {journal}
  {\bibinfo  {journal} {Phys. Rept.}\ }\textbf {\bibinfo {volume} {887}},\
  \bibinfo {pages} {1} (\bibinfo {year} {2020})},\ \Eprint
  {https://arxiv.org/abs/2006.04822} {arXiv:2006.04822 [hep-ph]} \BibitemShut
  {NoStop}%
\bibitem [{\citenamefont {Borsanyi}\ \emph {et~al.}(2021)\citenamefont
  {Borsanyi} \emph {et~al.}}]{Borsanyi:2020mff}%
  \BibitemOpen
  \bibfield  {author} {\bibinfo {author} {\bibfnamefont {S.}~\bibnamefont
  {Borsanyi}} \emph {et~al.},\ }\bibfield  {title} {\bibinfo {title} {{Leading
  hadronic contribution to the muon magnetic moment from lattice QCD}},\ }\href
  {https://doi.org/10.1038/s41586-021-03418-1} {\bibfield  {journal} {\bibinfo
  {journal} {Nature}\ }\textbf {\bibinfo {volume} {593}},\ \bibinfo {pages}
  {51} (\bibinfo {year} {2021})},\ \Eprint {https://arxiv.org/abs/2002.12347}
  {arXiv:2002.12347 [hep-lat]} \BibitemShut {NoStop}%
\bibitem [{\citenamefont {Davier}\ \emph {et~al.}(2011)\citenamefont {Davier},
  \citenamefont {Hoecker}, \citenamefont {Malaescu},\ and\ \citenamefont
  {Zhang}}]{Davier:2010nc}%
  \BibitemOpen
  \bibfield  {author} {\bibinfo {author} {\bibfnamefont {M.}~\bibnamefont
  {Davier}}, \bibinfo {author} {\bibfnamefont {A.}~\bibnamefont {Hoecker}},
  \bibinfo {author} {\bibfnamefont {B.}~\bibnamefont {Malaescu}},\ and\
  \bibinfo {author} {\bibfnamefont {Z.}~\bibnamefont {Zhang}},\ }\bibfield
  {title} {\bibinfo {title} {{Reevaluation of the Hadronic Contributions to the
  Muon $g-2$ and to $\alpha(M_Z)$}},\ }\href
  {https://doi.org/10.1140/epjc/s10052-012-1874-8} {\bibfield  {journal}
  {\bibinfo  {journal} {Eur. Phys. J. C}\ }\textbf {\bibinfo {volume} {71}},\
  \bibinfo {pages} {1515} (\bibinfo {year} {2011})},\ \bibinfo {note}
  {[Erratum: Eur.Phys.J.C 72, 1874 (2012)]},\ \Eprint
  {https://arxiv.org/abs/1010.4180} {arXiv:1010.4180 [hep-ph]} \BibitemShut
  {NoStop}%
\bibitem [{\citenamefont {Davier}\ \emph {et~al.}(2017)\citenamefont {Davier},
  \citenamefont {Hoecker}, \citenamefont {Malaescu},\ and\ \citenamefont
  {Zhang}}]{Davier:2017zfy}%
  \BibitemOpen
  \bibfield  {author} {\bibinfo {author} {\bibfnamefont {M.}~\bibnamefont
  {Davier}}, \bibinfo {author} {\bibfnamefont {A.}~\bibnamefont {Hoecker}},
  \bibinfo {author} {\bibfnamefont {B.}~\bibnamefont {Malaescu}},\ and\
  \bibinfo {author} {\bibfnamefont {Z.}~\bibnamefont {Zhang}},\ }\bibfield
  {title} {\bibinfo {title} {{Reevaluation of the hadronic vacuum polarisation
  contributions to the Standard Model predictions of the muon $g-2$ and
  ${\alpha (m_Z^2)}$ using newest hadronic cross-section data}},\ }\href
  {https://doi.org/10.1140/epjc/s10052-017-5161-6} {\bibfield  {journal}
  {\bibinfo  {journal} {Eur. Phys. J. C}\ }\textbf {\bibinfo {volume} {77}},\
  \bibinfo {pages} {827} (\bibinfo {year} {2017})},\ \Eprint
  {https://arxiv.org/abs/1706.09436} {arXiv:1706.09436 [hep-ph]} \BibitemShut
  {NoStop}%
\bibitem [{\citenamefont {Davier}\ \emph {et~al.}(2020)\citenamefont {Davier},
  \citenamefont {Hoecker}, \citenamefont {Malaescu},\ and\ \citenamefont
  {Zhang}}]{Davier:2019can}%
  \BibitemOpen
  \bibfield  {author} {\bibinfo {author} {\bibfnamefont {M.}~\bibnamefont
  {Davier}}, \bibinfo {author} {\bibfnamefont {A.}~\bibnamefont {Hoecker}},
  \bibinfo {author} {\bibfnamefont {B.}~\bibnamefont {Malaescu}},\ and\
  \bibinfo {author} {\bibfnamefont {Z.}~\bibnamefont {Zhang}},\ }\bibfield
  {title} {\bibinfo {title} {{A new evaluation of the hadronic vacuum
  polarisation contributions to the muon anomalous magnetic moment and to
  $\mathbf{\boldsymbol\alpha(m_Z^2)}$}},\ }\href
  {https://doi.org/10.1140/epjc/s10052-020-7792-2} {\bibfield  {journal}
  {\bibinfo  {journal} {Eur. Phys. J. C}\ }\textbf {\bibinfo {volume} {80}},\
  \bibinfo {pages} {241} (\bibinfo {year} {2020})},\ \bibinfo {note} {[Erratum:
  Eur.Phys.J.C 80, 410 (2020)]},\ \Eprint {https://arxiv.org/abs/1908.00921}
  {arXiv:1908.00921 [hep-ph]} \BibitemShut {NoStop}%
\bibitem [{\citenamefont {Keshavarzi}\ \emph {et~al.}(2020)\citenamefont
  {Keshavarzi}, \citenamefont {Marciano}, \citenamefont {Passera},\ and\
  \citenamefont {Sirlin}}]{Keshavarzi:2020bfy}%
  \BibitemOpen
  \bibfield  {author} {\bibinfo {author} {\bibfnamefont {A.}~\bibnamefont
  {Keshavarzi}}, \bibinfo {author} {\bibfnamefont {W.~J.}\ \bibnamefont
  {Marciano}}, \bibinfo {author} {\bibfnamefont {M.}~\bibnamefont {Passera}},\
  and\ \bibinfo {author} {\bibfnamefont {A.}~\bibnamefont {Sirlin}},\
  }\bibfield  {title} {\bibinfo {title} {{Muon $g-2$ and $\Delta \alpha$
  connection}},\ }\href {https://doi.org/10.1103/PhysRevD.102.033002}
  {\bibfield  {journal} {\bibinfo  {journal} {Phys. Rev. D}\ }\textbf {\bibinfo
  {volume} {102}},\ \bibinfo {pages} {033002} (\bibinfo {year} {2020})},\
  \Eprint {https://arxiv.org/abs/2006.12666} {arXiv:2006.12666 [hep-ph]}
  \BibitemShut {NoStop}%
\bibitem [{\citenamefont {Crivellin}\ \emph {et~al.}(2020)\citenamefont
  {Crivellin}, \citenamefont {Hoferichter}, \citenamefont {Manzari},\ and\
  \citenamefont {Montull}}]{Crivellin:2020zul}%
  \BibitemOpen
  \bibfield  {author} {\bibinfo {author} {\bibfnamefont {A.}~\bibnamefont
  {Crivellin}}, \bibinfo {author} {\bibfnamefont {M.}~\bibnamefont
  {Hoferichter}}, \bibinfo {author} {\bibfnamefont {C.~A.}\ \bibnamefont
  {Manzari}},\ and\ \bibinfo {author} {\bibfnamefont {M.}~\bibnamefont
  {Montull}},\ }\bibfield  {title} {\bibinfo {title} {{Hadronic Vacuum
  Polarization: $(g-2)_\mu$ versus Global Electroweak Fits}},\ }\href
  {https://doi.org/10.1103/PhysRevLett.125.091801} {\bibfield  {journal}
  {\bibinfo  {journal} {Phys. Rev. Lett.}\ }\textbf {\bibinfo {volume} {125}},\
  \bibinfo {pages} {091801} (\bibinfo {year} {2020})},\ \Eprint
  {https://arxiv.org/abs/2003.04886} {arXiv:2003.04886 [hep-ph]} \BibitemShut
  {NoStop}%
\bibitem [{\citenamefont {Isidori}\ \emph {et~al.}(2021)\citenamefont
  {Isidori}, \citenamefont {Lancierini}, \citenamefont {Owen},\ and\
  \citenamefont {Serra}}]{Isidori:2021vtc}%
  \BibitemOpen
  \bibfield  {author} {\bibinfo {author} {\bibfnamefont {G.}~\bibnamefont
  {Isidori}}, \bibinfo {author} {\bibfnamefont {D.}~\bibnamefont {Lancierini}},
  \bibinfo {author} {\bibfnamefont {P.}~\bibnamefont {Owen}},\ and\ \bibinfo
  {author} {\bibfnamefont {N.}~\bibnamefont {Serra}},\ }\bibfield  {title}
  {\bibinfo {title} {{On the significance of new physics in $b \rightarrow s
  \ell^+ \ell^-$ decays}},\ }\href
  {https://doi.org/10.1016/j.physletb.2021.136644} {\bibfield  {journal}
  {\bibinfo  {journal} {Phys. Lett. B}\ }\textbf {\bibinfo {volume} {822}},\
  \bibinfo {pages} {136644} (\bibinfo {year} {2021})},\ \Eprint
  {https://arxiv.org/abs/2104.05631} {arXiv:2104.05631 [hep-ph]} \BibitemShut
  {NoStop}%
\bibitem [{\citenamefont {Ciuchini}\ \emph {et~al.}(2021)\citenamefont
  {Ciuchini}, \citenamefont {Fedele}, \citenamefont {Franco}, \citenamefont
  {Paul}, \citenamefont {Silvestrini},\ and\ \citenamefont
  {Valli}}]{Ciuchini:2021smi}%
  \BibitemOpen
  \bibfield  {author} {\bibinfo {author} {\bibfnamefont {M.}~\bibnamefont
  {Ciuchini}}, \bibinfo {author} {\bibfnamefont {M.}~\bibnamefont {Fedele}},
  \bibinfo {author} {\bibfnamefont {E.}~\bibnamefont {Franco}}, \bibinfo
  {author} {\bibfnamefont {A.}~\bibnamefont {Paul}}, \bibinfo {author}
  {\bibfnamefont {L.}~\bibnamefont {Silvestrini}},\ and\ \bibinfo {author}
  {\bibfnamefont {M.}~\bibnamefont {Valli}},\ }\bibfield  {title} {\bibinfo
  {title} {{New Physics without bias: Charming Penguins and Lepton Universality
  Violation in $b \to s \ell^+ \ell^-$ decays}},\ }\href@noop {} {\  (\bibinfo
  {year} {2021})},\ \Eprint {https://arxiv.org/abs/2110.10126}
  {arXiv:2110.10126 [hep-ph]} \BibitemShut {NoStop}%
\bibitem [{\citenamefont {Cacciapaglia}\ \emph {et~al.}(2006)\citenamefont
  {Cacciapaglia}, \citenamefont {Csaki}, \citenamefont {Marandella},\ and\
  \citenamefont {Strumia}}]{Cacciapaglia:2006pk}%
  \BibitemOpen
  \bibfield  {author} {\bibinfo {author} {\bibfnamefont {G.}~\bibnamefont
  {Cacciapaglia}}, \bibinfo {author} {\bibfnamefont {C.}~\bibnamefont {Csaki}},
  \bibinfo {author} {\bibfnamefont {G.}~\bibnamefont {Marandella}},\ and\
  \bibinfo {author} {\bibfnamefont {A.}~\bibnamefont {Strumia}},\ }\bibfield
  {title} {\bibinfo {title} {{The Minimal Set of Electroweak Precision
  Parameters}},\ }\href {https://doi.org/10.1103/PhysRevD.74.033011} {\bibfield
   {journal} {\bibinfo  {journal} {Phys. Rev. D}\ }\textbf {\bibinfo {volume}
  {74}},\ \bibinfo {pages} {033011} (\bibinfo {year} {2006})},\ \Eprint
  {https://arxiv.org/abs/hep-ph/0604111} {arXiv:hep-ph/0604111} \BibitemShut
  {NoStop}%
\bibitem [{\citenamefont {Pomarol}\ and\ \citenamefont
  {Vega}(1994)}]{Pomarol:1993mu}%
  \BibitemOpen
  \bibfield  {author} {\bibinfo {author} {\bibfnamefont {A.}~\bibnamefont
  {Pomarol}}\ and\ \bibinfo {author} {\bibfnamefont {R.}~\bibnamefont {Vega}},\
  }\bibfield  {title} {\bibinfo {title} {{Constraints on CP violation in the
  Higgs sector from the rho parameter}},\ }\href
  {https://doi.org/10.1016/0550-3213(94)90611-4} {\bibfield  {journal}
  {\bibinfo  {journal} {Nucl. Phys. B}\ }\textbf {\bibinfo {volume} {413}},\
  \bibinfo {pages} {3} (\bibinfo {year} {1994})},\ \Eprint
  {https://arxiv.org/abs/hep-ph/9305272} {arXiv:hep-ph/9305272} \BibitemShut
  {NoStop}%
\bibitem [{\citenamefont {Gerard}\ and\ \citenamefont
  {Herquet}(2007)}]{Gerard:2007kn}%
  \BibitemOpen
  \bibfield  {author} {\bibinfo {author} {\bibfnamefont {J.~M.}\ \bibnamefont
  {Gerard}}\ and\ \bibinfo {author} {\bibfnamefont {M.}~\bibnamefont
  {Herquet}},\ }\bibfield  {title} {\bibinfo {title} {{A Twisted custodial
  symmetry in the two-Higgs-doublet model}},\ }\href
  {https://doi.org/10.1103/PhysRevLett.98.251802} {\bibfield  {journal}
  {\bibinfo  {journal} {Phys. Rev. Lett.}\ }\textbf {\bibinfo {volume} {98}},\
  \bibinfo {pages} {251802} (\bibinfo {year} {2007})},\ \Eprint
  {https://arxiv.org/abs/hep-ph/0703051} {arXiv:hep-ph/0703051} \BibitemShut
  {NoStop}%
\bibitem [{\citenamefont {Haber}\ and\ \citenamefont
  {O'Neil}(2011)}]{Haber:2010bw}%
  \BibitemOpen
  \bibfield  {author} {\bibinfo {author} {\bibfnamefont {H.~E.}\ \bibnamefont
  {Haber}}\ and\ \bibinfo {author} {\bibfnamefont {D.}~\bibnamefont {O'Neil}},\
  }\bibfield  {title} {\bibinfo {title} {{Basis-independent methods for the
  two-Higgs-doublet model III: The CP-conserving limit, custodial symmetry, and
  the oblique parameters S, T, U}},\ }\href
  {https://doi.org/10.1103/PhysRevD.83.055017} {\bibfield  {journal} {\bibinfo
  {journal} {Phys. Rev. D}\ }\textbf {\bibinfo {volume} {83}},\ \bibinfo
  {pages} {055017} (\bibinfo {year} {2011})},\ \Eprint
  {https://arxiv.org/abs/1011.6188} {arXiv:1011.6188 [hep-ph]} \BibitemShut
  {NoStop}%
\bibitem [{\citenamefont {Funk}\ \emph {et~al.}(2012)\citenamefont {Funk},
  \citenamefont {O'Neil},\ and\ \citenamefont {Winters}}]{Funk:2011ad}%
  \BibitemOpen
  \bibfield  {author} {\bibinfo {author} {\bibfnamefont {G.}~\bibnamefont
  {Funk}}, \bibinfo {author} {\bibfnamefont {D.}~\bibnamefont {O'Neil}},\ and\
  \bibinfo {author} {\bibfnamefont {R.~M.}\ \bibnamefont {Winters}},\
  }\bibfield  {title} {\bibinfo {title} {{What the Oblique Parameters $S$, $T$,
  and $U$ and Their Extensions Reveal About the 2HDM: A Numerical Analysis}},\
  }\href {https://doi.org/10.1142/S0217751X12500212} {\bibfield  {journal}
  {\bibinfo  {journal} {Int. J. Mod. Phys. A}\ }\textbf {\bibinfo {volume}
  {27}},\ \bibinfo {pages} {1250021} (\bibinfo {year} {2012})},\ \Eprint
  {https://arxiv.org/abs/1110.3812} {arXiv:1110.3812 [hep-ph]} \BibitemShut
  {NoStop}%
\bibitem [{\citenamefont {Dawson}\ and\ \citenamefont
  {Murphy}(2017)}]{Dawson:2017vgm}%
  \BibitemOpen
  \bibfield  {author} {\bibinfo {author} {\bibfnamefont {S.}~\bibnamefont
  {Dawson}}\ and\ \bibinfo {author} {\bibfnamefont {C.~W.}\ \bibnamefont
  {Murphy}},\ }\bibfield  {title} {\bibinfo {title} {{Standard Model EFT and
  Extended Scalar Sectors}},\ }\href
  {https://doi.org/10.1103/PhysRevD.96.015041} {\bibfield  {journal} {\bibinfo
  {journal} {Phys. Rev. D}\ }\textbf {\bibinfo {volume} {96}},\ \bibinfo
  {pages} {015041} (\bibinfo {year} {2017})},\ \Eprint
  {https://arxiv.org/abs/1704.07851} {arXiv:1704.07851 [hep-ph]} \BibitemShut
  {NoStop}%
\bibitem [{\citenamefont {Fan}\ \emph {et~al.}(2022)\citenamefont {Fan},
  \citenamefont {Tang}, \citenamefont {Tsai},\ and\ \citenamefont
  {Wu}}]{Fan:2022dck}%
  \BibitemOpen
  \bibfield  {author} {\bibinfo {author} {\bibfnamefont {Y.-Z.}\ \bibnamefont
  {Fan}}, \bibinfo {author} {\bibfnamefont {T.-P.}\ \bibnamefont {Tang}},
  \bibinfo {author} {\bibfnamefont {Y.-L.~S.}\ \bibnamefont {Tsai}},\ and\
  \bibinfo {author} {\bibfnamefont {L.}~\bibnamefont {Wu}},\ }\bibfield
  {title} {\bibinfo {title} {{Inert Higgs Dark Matter for New CDF W-boson Mass
  and Detection Prospects}},\ }\href@noop {} {\  (\bibinfo {year} {2022})},\
  \Eprint {https://arxiv.org/abs/2204.03693} {arXiv:2204.03693 [hep-ph]}
  \BibitemShut {NoStop}%
\bibitem [{\citenamefont {Lu}\ \emph {et~al.}(2022)\citenamefont {Lu},
  \citenamefont {Wu}, \citenamefont {Wu},\ and\ \citenamefont
  {Zhu}}]{Lu:2022bgw}%
  \BibitemOpen
  \bibfield  {author} {\bibinfo {author} {\bibfnamefont {C.-T.}\ \bibnamefont
  {Lu}}, \bibinfo {author} {\bibfnamefont {L.}~\bibnamefont {Wu}}, \bibinfo
  {author} {\bibfnamefont {Y.}~\bibnamefont {Wu}},\ and\ \bibinfo {author}
  {\bibfnamefont {B.}~\bibnamefont {Zhu}},\ }\bibfield  {title} {\bibinfo
  {title} {{Electroweak Precision Fit and New Physics in light of $W$ Boson
  Mass}},\ }\href@noop {} {\  (\bibinfo {year} {2022})},\ \Eprint
  {https://arxiv.org/abs/2204.03796} {arXiv:2204.03796 [hep-ph]} \BibitemShut
  {NoStop}%
\bibitem [{\citenamefont {Athron}\ \emph {et~al.}(2022)\citenamefont {Athron},
  \citenamefont {Fowlie}, \citenamefont {Lu}, \citenamefont {Wu}, \citenamefont
  {Wu},\ and\ \citenamefont {Zhu}}]{Athron:2022qpo}%
  \BibitemOpen
  \bibfield  {author} {\bibinfo {author} {\bibfnamefont {P.}~\bibnamefont
  {Athron}}, \bibinfo {author} {\bibfnamefont {A.}~\bibnamefont {Fowlie}},
  \bibinfo {author} {\bibfnamefont {C.-T.}\ \bibnamefont {Lu}}, \bibinfo
  {author} {\bibfnamefont {L.}~\bibnamefont {Wu}}, \bibinfo {author}
  {\bibfnamefont {Y.}~\bibnamefont {Wu}},\ and\ \bibinfo {author}
  {\bibfnamefont {B.}~\bibnamefont {Zhu}},\ }\bibfield  {title} {\bibinfo
  {title} {{The $W$ boson Mass and Muon $g-2$: Hadronic Uncertainties or New
  Physics?}},\ }\href@noop {} {\  (\bibinfo {year} {2022})},\ \Eprint
  {https://arxiv.org/abs/2204.03996} {arXiv:2204.03996 [hep-ph]} \BibitemShut
  {NoStop}%
\bibitem [{\citenamefont {Strumia}(2022)}]{Strumia:2022qkt}%
  \BibitemOpen
  \bibfield  {author} {\bibinfo {author} {\bibfnamefont {A.}~\bibnamefont
  {Strumia}},\ }\bibfield  {title} {\bibinfo {title} {{Interpreting electroweak
  precision data including the $W$-mass CDF anomaly}},\ }\href@noop {} {\
  (\bibinfo {year} {2022})},\ \Eprint {https://arxiv.org/abs/2204.04191}
  {arXiv:2204.04191 [hep-ph]} \BibitemShut {NoStop}%
\bibitem [{\citenamefont {de~Blas}\ \emph {et~al.}(2022)\citenamefont
  {de~Blas}, \citenamefont {Pierini}, \citenamefont {Reina},\ and\
  \citenamefont {Silvestrini}}]{blas2022impact}%
  \BibitemOpen
  \bibfield  {author} {\bibinfo {author} {\bibfnamefont {J.}~\bibnamefont
  {de~Blas}}, \bibinfo {author} {\bibfnamefont {M.}~\bibnamefont {Pierini}},
  \bibinfo {author} {\bibfnamefont {L.}~\bibnamefont {Reina}},\ and\ \bibinfo
  {author} {\bibfnamefont {L.}~\bibnamefont {Silvestrini}},\ }\bibfield
  {title} {\bibinfo {title} {Impact of the recent measurements of the top-quark
  and w-boson masses on electroweak precision fits},\ }\href@noop {} {\
  (\bibinfo {year} {2022})},\ \Eprint {https://arxiv.org/abs/2204.04204}
  {arXiv:2204.04204 [hep-ph]} \BibitemShut {NoStop}%
\bibitem [{\citenamefont {Kribs}\ \emph {et~al.}(2021)\citenamefont {Kribs},
  \citenamefont {Lu}, \citenamefont {Martin},\ and\ \citenamefont
  {Tong}}]{Kribs:2020jgn}%
  \BibitemOpen
  \bibfield  {author} {\bibinfo {author} {\bibfnamefont {G.~D.}\ \bibnamefont
  {Kribs}}, \bibinfo {author} {\bibfnamefont {X.}~\bibnamefont {Lu}}, \bibinfo
  {author} {\bibfnamefont {A.}~\bibnamefont {Martin}},\ and\ \bibinfo {author}
  {\bibfnamefont {T.}~\bibnamefont {Tong}},\ }\bibfield  {title} {\bibinfo
  {title} {{Custodial symmetry violation in the SMEFT}},\ }\href
  {https://doi.org/10.1103/PhysRevD.104.056006} {\bibfield  {journal} {\bibinfo
   {journal} {Phys. Rev. D}\ }\textbf {\bibinfo {volume} {104}},\ \bibinfo
  {pages} {056006} (\bibinfo {year} {2021})},\ \Eprint
  {https://arxiv.org/abs/2009.10725} {arXiv:2009.10725 [hep-ph]} \BibitemShut
  {NoStop}%
\end{thebibliography}%

\end{document}